\documentclass[sigconf, natbib]{acmart}

\usepackage[utf8]{inputenc}
\usepackage{url}
\usepackage{listings}
\usepackage{subcaption}
\usepackage{bbding}
\usepackage{pifont}
\usepackage{array}

\usepackage{caption}
 
\usepackage{newfloat}
\DeclareFloatingEnvironment[
   fileext=lst,
   name=Listing
]{listing}
\floatname{listing}{Listing}

\newcommand\code[1]{{\small{\texttt{#1}}}}
\newcommand\fncode[1]{{\scriptsize{\texttt{#1}}}}

\usepackage{xspace}
\usepackage{etoolbox} 
\newcommand{\CC}{C\nolinebreak\hspace{-.05em}\raisebox{.2ex}{\small +}\nolinebreak\hspace{-.05em}\raisebox{.2ex}{\small +}\xspace}

\robustify{\CC}
\newcommand{\ompi}{Open~MPI\xspace}

\newcommand{\rulesep}{\unskip\ \vrule\ }

\title{Fibers are not (P)Threads}
\subtitle{The Case for Loose Coupling of Asynchronous Programming Models and MPI Through Continuations}
\copyrightyear{2020}
\acmYear{2020}
\setcopyright{acmlicensed}\acmConference[EuroMPI/USA '20]{27th European MPI Users' Group Meeting}{September 21--24, 2020}{Austin, TX, USA}
\acmBooktitle{27th European MPI Users' Group Meeting (EuroMPI/USA '20), September 21--24, 2020, Austin, TX, USA}
\acmPrice{15.00}
\acmDOI{10.1145/3416315.3416320}
\acmISBN{978-1-4503-8880-1/20/09}

\author{Joseph Schuchart}
\email{schuchart@hlrs.de}
\affiliation{
  \institution{Höchstleistungsrechenzentrum Stuttgart (HLRS)}
  \streetaddress{Nobelstraße 19}
  \postcode{70597}
  \city{Stuttgart}
  \country{Germany}
}
\author{Christoph Niethammer}
\email{niethammer@hlrs.de}
\affiliation{
  \institution{Höchstleistungsrechenzentrum Stuttgart (HLRS)}
  \streetaddress{Nobelstraße 19}
  \postcode{70597}
  \city{Stuttgart}
  \country{Germany}
}

\author{Jos\'{e} Gracia}
\email{gracia@hlrs.de}
\affiliation{
  \institution{Höchstleistungsrechenzentrum Stuttgart (HLRS)}
  \streetaddress{Nobelstraße 19}
  \postcode{70597}
  \city{Stuttgart}
  \country{Germany}
}


\graphicspath{{figures/}}

\begin{document}

\begin{abstract}
Asynchronous programming models (APM) are gaining more and more traction, allowing applications to expose the available concurrency to a runtime system tasked with coordinating the execution.
While MPI has long provided support for multi-threaded communication and non-blocking operations, it falls short of adequately supporting APMs as correctly and efficiently handling MPI communication in different models is still a challenge.
Meanwhile, new low-level implementations of light-weight, cooperatively scheduled execution contexts (\emph{fibers}, aka \emph{user-level threads} (ULT)) are meant to serve as a basis for higher-level APMs and their integration in MPI implementations has been proposed as a replacement for traditional POSIX thread support to alleviate these challenges.

In this paper, we first establish a taxonomy in an attempt to clearly distinguish different concepts in the parallel software stack.
We argue that the proposed tight integration of fiber implementations with MPI is neither warranted nor beneficial and instead is detrimental to the goal of MPI being a portable communication abstraction.
We propose \emph{MPI Continuations} as an extension to the MPI standard to provide callback-based notifications on completed operations, leading to a clear separation of concerns by providing a loose coupling mechanism between MPI and APMs.
We show that this interface is flexible and interacts well with different APMs, namely OpenMP detached tasks, OmpSs-2, and Argobots.
\end{abstract}
\maketitle

\section{Background and Motivation}

Asynchronous (task-based) programming models have gained more and more traction, promising to help users better utilize ubiquitous multi- and many-core systems by exposing all available concurrency to a runtime system.
Applications are expressed in terms of work-packages (tasks) with well-defined inputs and outputs, which guide the runtime scheduler in determining the correct execution order of tasks.
A wide range of task-based programming models have been developed, ranging from node-local approaches such as OpenMP~\cite{openmp5.0} and OmpSs~\cite{ompss2011} to distributed systems such as HPX~\cite{hpx:2014}, StarPU~\cite{Augonnet:2011}, DASH~\cite{Schuchart:2019:GTD}, and PaRSEC~\cite{DAGuE:2011}.
All of them have in common that a set of tasks is scheduled for execution by a set of worker threads based on constraints provided by the user, either in the form of dependencies or dataflow expressions.

At the same time, MPI is still the dominant interface for inter-process communication in parallel applications~\cite{Berholdt:2018:MPIUsage}, providing blocking and non-blocking point-to-point and collective operations as well as I/O capabilities on top of elaborate datatype and process group abstractions~\cite{mpi3.1}.
Recent years have seen significant improvements to the way MPI implementations handle communication by multiple threads of execution in parallel~\cite{Patinyasakdikul:2019:GMT,Hjelm:2018:IMM}.

However, task-based programming models pose new challenges to the way applications interact with MPI.
While MPI provides non-blocking communications that can be used to hide communication latencies, it is left to the application layer to ensure that all necessary communication operations are eventually initiated to avoid deadlocks, that in-flight communication operations are eventually completed, and that the interactions between communicating tasks are correctly handled.
Higher-level tasking approaches such as PaRSEC or HPX commonly track the state of active communication operations and regularly test for completion before acting upon such a change in state.
With node-local programming models such as OpenMP, this tracking of active communication operations is left to the application layer.
Unfortunately, the simple approach of test-yield cycles inside a task does not provide optimal performance due to CPU cycles being wasted on testing and (in the case of OpenMP) may not even be portable~\cite{Schuchart:2018:TIT}.

MPI is thus currently not well equipped to support users and developers of asynchronous programming models, which (among other things) has prompted some higher level runtimes to move away from MPI towards more asynchronous APIs~\cite{Daliss:2019:FPD}.

Different approaches have been proposed to mitigate this burden, including Task-Aware MPI (TAMPI)~\cite{Sala:2019:TAMPI} and a \emph{tight integration} of user-level thread, fiber, and tasking libraries (cf. \autoref{sec:taxonomy}) with MPI implementations~\cite{Mercier:2009:NMA,Lu:2015:MUO,Sala:2018:IIM}.
In this paper, we argue that the latter is the least preferable solution and in fact should be seen critical by the MPI community.\footnote{This paper is in part an extended version of an argument made online at \url{https://github.com/open-mpi/ompi/pull/6578\#issuecomment-602542762}.}
After establishing a taxonomy to be used throughout this paper (\autoref{sec:taxonomy}), we will make the argument that such an integration is neither warranted by the MPI standard nor beneficial to the goal of supporting the development of portable MPI applications (\autoref{sec:aspects}).
Instead, we propose adding \emph{continuations} to the MPI standard to facilitate a \emph{loose coupling} between higher-level concurrency abstractions and MPI (\autoref{sec:continuations}).
We will show that the interface is flexible and can be used with abstractions on different levels, including OpenMP detached tasks, OmpSs-2 task events, and Argobots (\autoref{sec:evaluation}).

\section{Taxonomy}
\label{sec:taxonomy}

\begin{figure}
\centering
\includegraphics[width=.8\columnwidth]{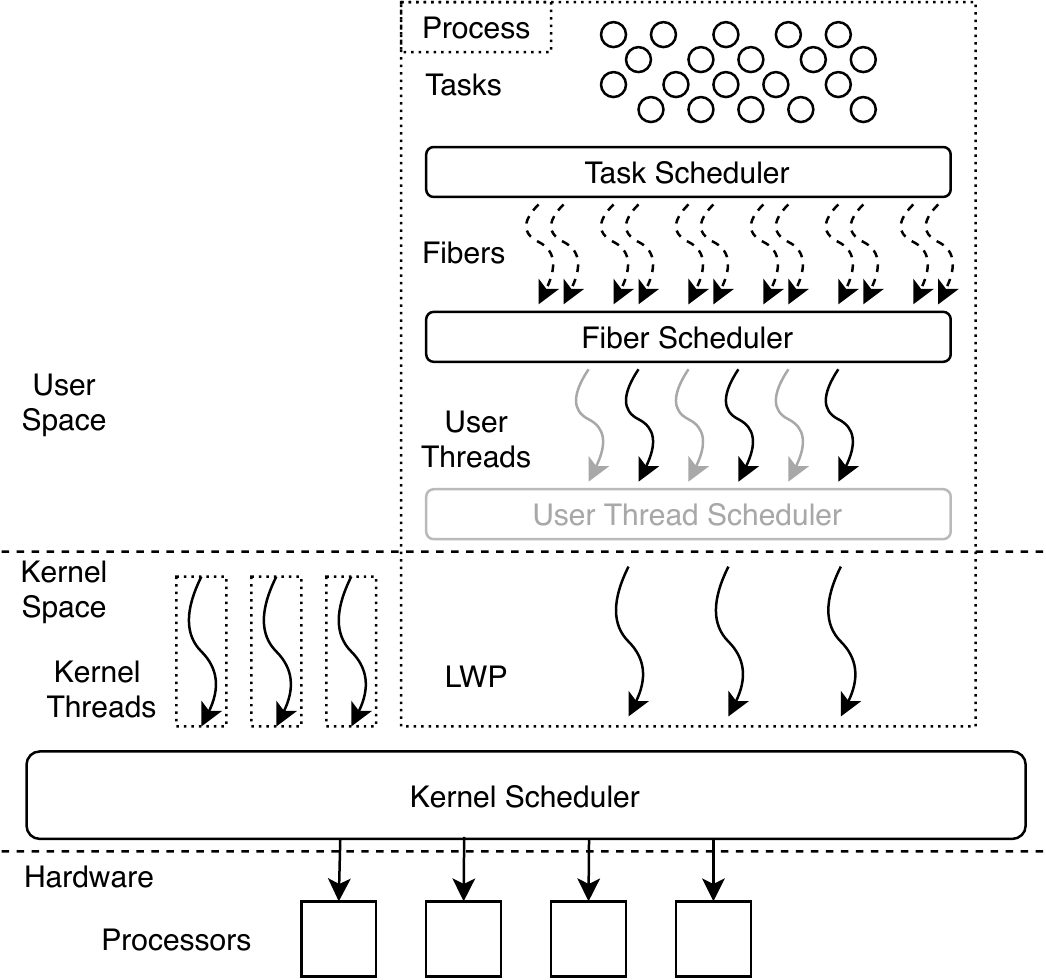}
\caption{Hierarchy of kernel and user threads, fibers, and tasks with the different scheduler entities. The user thread scheduler is optional and not present in most modern architectures. Tasks may be executed directly on top of threads without the use of fibers.}
\label{fig:thread_architecture}
\end{figure}

Unfortunately, the term \emph{thread} (or \emph{thread of execution}) is not well-defined in the computer science literature.
Terms like \emph{kernel threads}, \emph{(light-weight) processes}, \emph{user-level} and \emph{user-space threads}, \emph{POSIX threads} (pthreads), \emph{light-weight threads}, and \emph{fibers} are used without a clear consensus on a common taxonomy.
Establishing such a taxonomy is thus warranted to better guide the discussion in the remainder of this paper.
Some of the description is based on the terminology provided in~\cite{UNIXInternals} and~\cite{Silberschatz:2014:OSC}.
The relation between the concepts is depicted in \autoref{fig:thread_architecture}, described here from bottom to top.

\paragraph{Processor}
A processor provides the active physical resources to execute a stream of instructions.
In the context of this work, the term referes to a unicore CPU or a single core on a multi-core CPU.

\paragraph{Kernel Thread}
A kernel thread is a thread of execution that is entirely controlled by the operating system kernel.
Kernel threads only execute within the kernel space and serve tasks such as asynchronous I/O, memory management, and signal handling.
They also form the basis of light-weight processes.

\paragraph{Process}
A process is an execution instance of a program and a user-level abstraction of state, providing isolation of kernel-level resources such as memory pages and file pointers between different processes.
A process may use one or more light-weight processes.

\paragraph{Light-weight Process (LWP)}
A light-weight process can be considered an abstraction of a processor inside the operating system (OS), on which threads are scheduled.
LWPs are scheduled preemptively by the OS kernel, i.e., their execution may be preempted and another LWP be scheduled onto the used processor.
Different LWPs may share resources such as file descriptors and memory if they belong to the same process.

\paragraph{Cooperative and Preemptive Scheduling}
These are two main scheduling strategies to distinguish.
With \emph{cooperative scheduling}, a scheduler assigns resources to a process or thread until it relinquishes them.
The benefits of cooperative scheduling are lower scheduling overheads (scheduling decisions are made only at well-defined points) and---at least on single-core systems---atomicity of execution between scheduling points~\cite{Adya:2002:CTM}.
With \emph{preemptive scheduling}, the scheduler is able to preempt the execution of a thread after the expiration of a time-slice or to schedule processes or threads with higher priority.
Most modern OS kernels employ preemptive scheduling to ensure fairness and responsiveness.

\paragraph{User Thread}%

\emph{User threads} are threads of execution that share resources within the confines of a process.
$N$ threads may be statically assigned to $N$ LWP (1:1 mapping) or multiplexed onto either a single (1:N mapping) or $M$ LWPs (N:M mapping) by a user thread scheduler.
In all cases, threads are comprised of some private state (stack memory, register context, thread-local variables, signal masks, \code{errno} value, scheduling priority and policy) and have access to resources shared between threads within a process (heap memory, global variables, file descriptors, etc).

A common example of a \emph{user thread} implementation are POSIX threads (\emph{pthreads})~\cite{POSIX2018}.
In all currently available Unix flavors known to the authors, a 1:1 mapping between pthreads and LWPs is maintained, including Linux~\cite{Pthread:Linux} and various BSD flavors~\cite{Pthread:FreeBDS,Pthread:OpenBSD,Pthread:NetBSD}.
They are the basic building block for any type of multi-threaded applications and runtime systems.
Solaris  provided its own thread implementation (Solaris threads), which originally employed an N:M mapping that was abandoned eventually for efficiency reasons~\cite{SolarisMT}, as were previous N:M implementation in BSD.
POSIX, Solaris, and Windows threads~\cite{MSThreads} are scheduled preemptively.\footnote{We note that in POSIX.1, the default scheduler \fncode{SCHED\_OTHER} does not mandate preemptive scheduling of threads but all implementations (both N:M and 1:1) known to the authors are scheduled preemptively. Applications and libraries relying on preemptive scheduling for correctness are thus \emph{de-facto} correct. \emph{De-jure} correct applications are required to periodically call \fncode{sched\_yield} or use pthread-specific synchronization mechanisms to ensure progress.}
We will refer to \emph{user threads} as defined here simply as \emph{threads} in the remainder of this work.

\paragraph{Fiber}
Fibers differ from threads in that multiple fibers share some of the context of the thread they are executing on, i.e., they are not fully independent execution entities~\cite{Goodspeed:2014:DCF}.
A fiber is comprised of a private stack and a memory region to save relevant registers when the fiber is rescheduled.
However, fibers executing on the same thread have access to the same thread-private variables and share the priority and time-slice of the thread.
They are unknown to the OS, i.e., they do not posses a thread context assigned by the OS or user thread scheduler, and are cooperatively scheduled for execution on threads by simply changing the stack pointer and adjusting the register context.
A fiber blocked in a library call will block the thread until the operation has completed.
A fiber blocked in a system call will cause the executing thread to be blocked (and potentially rescheduled in lieu of another thread).

Fiber implementations may provide synchronization primitives that reschedule the fiber to allow the underlying thread to execute a different fiber if the execution would otherwise be blocked.
However, third-party libraries have to actively support a particular fiber implementation.
While fibers are a way to express \emph{concurrency}, their use does not impact the \emph{parallelism} of an application, which is a function of the number of available processors and the number of threads used to execute fibers.

A wide range of fiber implementations is available:  in Microsoft Windows~\cite{MSFibers}, the Boost.Fiber \CC library~\cite{BoostFiber}, the GNU pth package~\cite{GNUPth}\footnote{Note that the GNU Pth author distinguishes between \emph{kernel-space} and \emph{user-space} threads to distinguish 1:1 and 1:N mappings between threads and LWPs and mentions the possibility of hybrid forms. We feel that this distinction is confusing  with regards to the existence of \emph{kernel threads} described above.}, a custom implementations in OmpSs~\cite{ompss2011}, Qthreads~\cite{Wheeler:2008:QAA}, as well as the more recent Argobots~\cite{Seo:2018:AAL}.
The POSIX \code{makecontext} set of functions can be used to implement fibers~\cite{Makecontext:Linux}.

The main difference between fibers and threads should be seen in that a thread is \emph{the lowest executing entity in user space} while a fiber may provide some execution context and isolation but \emph{always requires an underlying thread} for execution.

In the literature, fibers are also referred to as \emph{lightweight threads} and \emph{user-level threads}~\cite{Wheeler:2008:QAA,Seo:2018:AAL}.
However, we believe that the term \emph{thread} (esp. in light of the ensuing discussion) is convoluted and a clear distinction between threads and fibers is necessary.

\paragraph{Task}
A task is a unit of work that is scheduled for execution by a task scheduler, potentially based on constraint (e.g., dependencies) provided by the user.
The task scheduler maps the execution of a set of tasks to a set of worker threads, which may execute the associated action either directly on their stack or encapsulate the execution within a fiber to allow for rescheduling of executing tasks.
Note that the task and fiber schedulers depicted in \autoref{fig:thread_architecture} may be collapsed into a single instance managing both aspects.

\section{The Case Against A Tight Fiber Integration In MPI}
\label{sec:aspects}

In the discussion of whether fiber implementations should be tightly integrated with MPI implementations, we will look at three aspects: i)~\emph{correctness} (of the MPI implementation and the application); ii)~\emph{portability}; and iii)~potential \emph{performance} issues.

\subsection{Correctness}
\label{sec:correctness}
When considering correctness in the context of thread-parallel MPI it is important to differentiate between \emph{MPI-level correctness} and \emph{application-level correctness}.

The MPI standard mandates that \emph{thread-compliant} implementations should a) be \emph{thread-safe}; and b) only block the calling thread of a ``thread package similar to POSIX threads''~\cite[\S12.4]{mpi3.1}.
Thread-safety requires MPI implementations to protect any internal data structures from corruption in case multiple threads make calls into the MPI library in parallel, either calling into the same or different procedures.
This is commonly achieved through mutual exclusion devices provided by the underlying threading library, e.g., using instances of pthread mutex, or by using operations to atomically modify values in memory, which are provided by the processor.

More interesting, however, is the second requirement: blocking operations should only block the calling thread and allow other threads to continue progressing their operations.
This is especially important in cases where the successful completion of one thread's operation depends on the completion of another thread's operations, either directly (thread A sending a message to thread B on the same process) or indirectly (thread $A_{p_0}$ sending a message to thread $A_{p_1}$ on a different process, which then responds with a message that thread $B_{p_0}$ is waiting on).
In the context of POSIX threads, this can be achieved by reducing the time mutexes are held.

Part of the correctness requirement of a thread-compliant MPI implementation is the avoidance of deadlocks at the \emph{implementation level}, e.g., by avoiding lock order inversion.
However, MPI implementations are not required to ensure that all threads are scheduled for execution to avoid deadlocks at the \emph{application level} since they are scheduled preemptively: once a thread blocked in an MPI call exceeds its time-slice it will be replaced by another thread that may then initiate its communication.
Eventually, all threads will have initiated their communication and the application progresses.

\begin{listing}
\begin{lstlisting}
int myrank;
void* threadfn(intptr_t tag) {
  int val = 0;
  MPI_Recv(&val, 1, MPI_INT, myrank, 
          tag, MPI_COMM_WORLD, MPI_STATUS_IGNORE);
  return NULL:
}
void test() {
  MPI_Comm_rank(MPI_COMM_WORLD, &myrank);
  pthread_t threads[NUM_THREADS];
  /* create receiving threads */
  for (intptr_t i = 0; i < NUM_THREADS; ++i) {
    pthread_create(&threads[i], NULL, &threadfn, (void*)i);
  }
  /* send their messages */
  for (int i = 0; i < NUM_THREADS; ++i) {
    int val;
    MPI_Send(&val, 1, MPI_INT, myrank, i, MPI_COMM_WORLD);
    pthread_join(threads[i], NULL);
  }
}
\end{lstlisting}
\caption{Example of multiple threads receiving messages from a single thread on the same process.}
\label{lst:pthread_send_recv}
\end{listing}

\begin{figure}
\begin{subfigure}{.32\columnwidth}
\centering
\includegraphics[width=.9\textwidth, page=5]{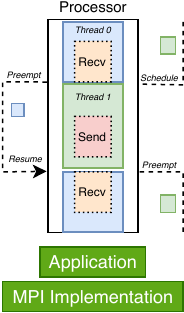}
\subcaption{Preemptively scheduled \emph{threads}}
\label{fig:correctness:threads}
\end{subfigure}
\rulesep
\begin{subfigure}{.32\columnwidth}
\centering
\includegraphics[width=.9\textwidth, page=7]{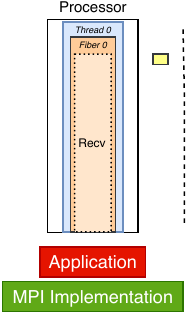}
\subcaption{\emph{Fibers} (without support in MPI)}
\label{fig:correctness:nofib}
\end{subfigure}
\rulesep
\begin{subfigure}{.32\columnwidth}
\centering
\includegraphics[width=.9\textwidth, page=6]{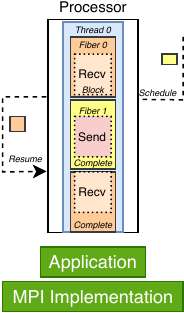}
\subcaption{\emph{Fibers} (with support in MPI)}
\label{fig:correctness:fib}
\end{subfigure}
\caption{Correctness of threads and fibers communicating through MPI, with and without fiber integration.}
\label{fig:correctness}
\end{figure}

\autoref{lst:pthread_send_recv} provides an example of multiple threads being spawned waiting to receive a message through MPI from the main thread.
A schematic depiction of parts of the execution is shown in \autoref{fig:correctness:threads}.
This example will always succeed, even for large numbers of receiving threads, due to the preemptive scheduling of POSIX thread implementations.
It is thus a correct MPI program.

MPI implementations may employ thread synchronization mechanisms to improve the \emph{efficiency} of multi-threaded MPI applications but such optimizations are not required for \emph{correctness}.
In \autoref{lst:pthread_send_recv}, we can replace the used blocking with non-blocking MPI procedures and a tight test-loop to prevent such optimizations inside the MPI implementation.
The result will be the same.

Replacing the calls to \code{pthread\_create} in \autoref{lst:pthread_send_recv} with the respective calls in a fiber library, however, would not yield a correct program even if the MPI implementation called \code{sched\_yield}.
Depending on the scheduling order of the fibers, the application may deadlock because fibers may be executed for which the main thread does not currently send a message.
The application has to be considered erroneous if it cannot guarantee that all relevant communication operations are initiated.\footnote{Such behavior would be comparable with Example~3.8 of the MPI standard, which calls an attempt to use blocking receive calls on two communicating processes before initiating the send operations \emph{errant}~\cite[p.~43]{mpi3.1}.}
Such an erroneous execution is depicted in \autoref{fig:correctness:nofib}, where the thread executing the receiving fiber (orange) is blocked indefinitely inside MPI and the sending fiber (yellow) is never scheduled for execution.

It should be noted that the behavior of the MPI implementation in that case is perfectly legitimate: since fibers are executed by threads, the implementation may use mutexes to guard its internal data structures against corruption from two or more \emph{threads} making MPI calls in parallel.
The correctness of the application's communication pattern is beyond the scope of the MPI implementation.

Correctness may be the main argument for a tight integration of (some) fiber libraries into MPI implementations to ease the burden for users.
However, the MPI standard does not mandate MPI implementations to support threading libraries beyond the basic building blocks (such as pthreads).
Thus, having two implementations---one with basic pthread support (\autoref{fig:correctness:nofib}) and one with support for the used fiber library (\autoref{fig:correctness:fib})---results in a situation in which both libraries are correct (from the MPI standard's viewpoint) but only one produces a correct outcome.
Hence, applications using blocking MPI communication inside fibers have to rely on an \emph{implementation detail} to ensure correct execution.
This raises significant concerns regarding the portability of MPI applications employing fibers.

\subsection{Portability}

Portability is a major goal of the MPI standard, which mentions the term \emph{portable} at least 88 times---once every ten pages---and explicitly states that ''the interface should establish a practical, portable, efficient, and flexible standard for message-passing.``~\cite{mpi3.1}
It is thus important for any solution tackling the issue of task-parallel MPI communication to provide means that are portable across multiple implementations.

\autoref{lst:omp_send_recv} shows a version of the example provided in \autoref{lst:pthread_send_recv} using OpenMP tasks.
As soon as \code{NUM\_TASKS} exceeds the number of threads available in the OpenMP \code{parallel} section this application may deadlock with most combinations of MPI+OpenMP implementations.
\autoref{tab:mpi_omp_compat} provides an overview of such combinations under which this application may run to completion for arbitrary \code{NUM\_TASKS}, i.e., won't deadlock.
Note that the positive combination of BOLT~\cite{Iwasaki:2019:BOLT} and \ompi/MPICH~\cite{Lu:2015:MUO} is not unconditional: at the time of this writing, the support for Argobots (and Qthreads) in both MPI implementations depends on a compile-time configuration switch of the MPI library. 
Applications would thus have to rely on either building their own MPI library or on the respective system administrators to provide---and on users to choose---the installation \emph{with support for the correct fiber library enabled}.
A major issue for applications relying on the availability of fiber support in MPI libraries is that this information is not exposed to the application layer, i.e., an application has no way of knowing whether the underlying MPI library was configured appropriately.

\begin{listing}
\begin{lstlisting}
  /* create receiving tasks */
  for (int i = 0; i < NUM_TASKS; ++i) {
    #pragma omp task
    {
      MPI_Recv(&buf[i], 1, MPI_INT, myrank, 
               i, MPI_COMM_WORLD, MPI_STATUS_IGNORE);
    }
  }
  /* send the messages */
  for (int i = 0; i < NUM_TASKS; ++i) {
    MPI_Send(&i, 1, MPI_INT, myrank, i, MPI_COMM_WORLD);
  }
\end{lstlisting}
\caption{Example of multiple \emph{OpenMP tasks} receiving messages from the master thread on the same process.}
\label{lst:omp_send_recv}
\end{listing}

By offering such integration in releases of mainstream MPI implementations, the community is actively inviting application developers to start relying on such tight integration, fostering the development of \emph{non-portable MPI applications}.
The resulting incompatibilities and efforts for debugging deadlocks will affect a wide range of the MPI community.
If, however, users were not encouraged to actually use these features the whole effort would be moot and the integration---including the disruptions to the threading support that is part of an implementation's core, the required testing, and efforts for integration into additional third-party libraries---was entirely unnecessary.
Either way, the efforts to integrate fiber models into mainstream releases of MPI libraries should be considered harmful.

\begin{table}
\caption{%
MPI/OpenMP task support matrix for using blocking MPI communication inside tasks. %
}
\label{tab:mpi_omp_compat}

\newcommand{\Yes}{\Checkmark}
\newcommand{\xmark}{\ding{55}}
\newcommand{\No}{\xmark}

\begin{tabular}{lccccc}
\toprule
\textbf{MPI / OpenMP} & \textbf{BOLT} & \textbf{GCC} & \textbf{Clang} & \textbf{Intel} & \textbf{Cray} \\
\midrule
Open MPI master & (\Yes)  & \No   & \No     & \No     &  \No \\
Open MPI <5.0 & \No    & \No   & \No     & \No     &  \No \\
MPICH        & (\Yes)   & \No   & \No     & \No     &  \No \\
MVAPICH      & \No    & \No   & \No     & \No     &  \No \\
Intel MPI    & \No    & \No   & \No     & \No     &  \No \\
Cray MPICH   & \No    & \No   & \No     & \No     &  \No \\
Bull MPI     & \No    & \No   & \No     & \No     &  \No \\
SGI/HP MPT & \No    & \No   & \No     & \No     &  \No \\ 
\bottomrule
\end{tabular}
\end{table}

\subsection{Portable Performance}
One of the major differences in the use of threads and fibers is \emph{oversubscription}: while applications and runtime systems typically try to avoid the use of more worker threads than available processors to mitigate the negative consequences of kernel-level scheduling, oversubscription is at the heart of the use of fibers.
Thanks to the low-overhead switching between fibers entirely in user-space, it is easily possible to create more fibers than worker threads and hide I/O-related latencies, including communication and file I/O.

However, if low latencies in the execution of a sequence of operations are paramount to the overall performance of an algorithm, unsolicited context switching inside MPI may negatively impact overall application performance.
As an example, application-level communication protocols using MPI RMA may rely on low-latency accumulate operations~\cite{Schuchart:2019:UMR}.
Any delay may result in unnecessarily increased latencies with the application layer having no way to control the context switching initiated by the MPI implementation.
While many communication latencies should be hidden through fiber oversubscription, some can be tolerated and it should be up to the application to decide between the two cases.

\section{MPI Continuations}
\label{sec:continuations}
Instead of tightly integrating various asynchronous execution libraries, we advocate for a mechanism to accomplish \emph{loose coupling} between MPI and the higher-level concurrency abstractions.

Continuations are a concept for structuring the execution of different parts of an application's code, dating back to research on Algol60~\cite{Friedman:1984:PWC,Reynolds;1993:TDC}.
They have recently been proposed to the \CC standard in the form of \code{std::future::then} to coordinate asynchronous activities~\cite{CPP:2014:stdfuture}.
Continuations consist of a \emph{body} (the code to execute) and a \code{context} (some state passed to the body) on which to operate.

A similar mechanism for MPI can be devised that allows a callback to be attached to an MPI request, which will be invoked once the operation represented by the request has completed.
This establishes a notification scheme to be used by applications to timely react to the completion of operations, e.g., to wake up a blocked fiber or release the dependencies of a detached task, all while relieving applications of managing MPI request objects themselves.

We propose a set of functions to set up \emph{continuations} for active MPI operations, which may be invoked by application threads during the execution of communication-related MPI functions or by an implementation-internal progress thread (if available) once all relevant operations are found to have completed.
The body of a continuation may call any MPI library function and thus start new MPI operations in response to the completion of previous ones.

The MPI Continuations API consists of two parts.
First, \emph{continuation requests} are used to aggregate and progress continuations.
Second, continuations are then attached to active MPI operations and registered with continuation requests for tracking.

\begin{listing}
\begin{lstlisting}
/* The continuation callback function definition */
typedef void (MPIX_Continue_cb_function)(MPI_Status *statuses,
                                         void       *cb_data);

/* Create a continuation request. */
int MPIX_Continue_init(MPI_Request  *cont_req);
  
/* Attach a continuation to an active operation represented by
 * op_request. Upon completion of the operation, the callback
 * cb will be invoked and passed status and cb_data as arguments.
 * The status object will be set before the continuation is invoked.
 * If the operation has completed already the continuation will
 * not be attached or invoked and flag will be set to 1. */
int MPIX_Continue(
  MPI_Request               *op_request,
  int                       *flag,
  MPIX_Continue_cb_function *cb,
  void                      *cb_data,
  MPI_Status                *status,
  MPI_Request                cont_req);
  
/* Similar to the above except that the continuation is only
 * invoked once all op_requests have completed. */
int MPIX_Continueall(
  int                        count,
  MPI_Request                op_requests[],
  int                       *flag,
  MPIX_Continue_cb_function *cb,
  void                      *cb_data,
  MPI_Status                 statuses[],
  MPI_Request                cont_req);
\end{lstlisting}
\caption{MPI Continuation interface.}
\label{lst:mpi_continue_fn}
\end{listing}

\subsection{Continuation Requests}

\emph{Continuation requests} (CR) are a form of persistent requests that are created through a call to \code{MPIX\_Continue\_init} and released eventually using \code{MPI\_Request\_free}.
A CR tracks a set of active continuations that are registered with it.
The set grows upon registration of a new continuation and shrinks once a continuation has been executed.
A call to \code{MPI\_Test} on a CR returns \code{flag\,==\,1} if no active continuations are registered.
Conversely, a call to \code{MPI\_Wait} blocks until all registered continuations have completed.
Further details are discussed in \autoref{sec:rationale}.

Continuation requests serve two main purposes.
First, they aggregate continuations attached to active operations and enable testing and waiting for their completion.
Second, by calling \code{MPI\_Test} on a CR, applications are able to progress outstanding operations and continuations if no MPI-internal progress  mechanism exists to process them.
See \autoref{sec:rationale_progress} for a discussion on progress.

\subsection{Registration of Continuations}
A continuation is attached to one or several active operations and registered with the \emph{continuation request} for tracking.
A call to \code{MPIX\_Continue} attaches a continuation to a single operation request while the use of \code{MPIX\_Continueall} sets up the continuation to be invoked once \emph{all} operations represented by the provided requests have completed.
As shown in \autoref{lst:mpi_continue_fn}, the continuation is represented through a callback function with the signature of \code{MPIX\_Continue\_cb\_function} (provided as function pointer \code{cb}) and a context (\code{cb\_data}).
Together with the pointer value provided for the \code{status} or \code{statuses} argument, the \code{cb\_data} will be passed to \code{cb} upon invocation.

Upon return, all provided non-persistent requests are set to \code{MPI\_REQUEST\_NULL}, effectively releasing them back to MPI.\footnote{This has been a deliberate design decision. Otherwise, the release of a request object inside the continuation or the MPI library would have to be synchronized with operations on it outside of the continuation. We thus avoid this source of errors.}
No copy of the requests should be used to cancel or test/wait for the completion of the associated operations.
Consequently, only one continuation may be attached to a non-persistent operation.
In contrast, persistent requests may still be canceled, tested, and waited for.
More than one continuation may be attached to a persistent request but the continuations themselves are not persistent.

Similar to \code{MPI\_Test}, \code{flag} shall be set to \code{1} if all operations have completed already.
In that case the continuation callback shall \emph{not} be invoked by MPI, leaving it to the application to handle immediate completion (see \autoref{sec:restrictions}).
Otherwise, \code{flag} is set to zero.

Unless  \code{MPI\_STATUS\_IGNORE} or \code{MPI\_STATUSES\_IGNORE} has been provided, the status object(s) will be set before the continuation is invoked (or before returning from \code{MPIX\_Continue[all]} in case all operations have completed already).
The status objects should be allocated by the application in a memory location that remains valid until the operation has completed and the callback has been invoked, e.g., on the stack of a blocked fiber or in memory allocated on the heap that is released inside the continuation callback.

\subsection{Control Flow Example}

\begin{figure}
\centering
\includegraphics[width=.75\columnwidth]{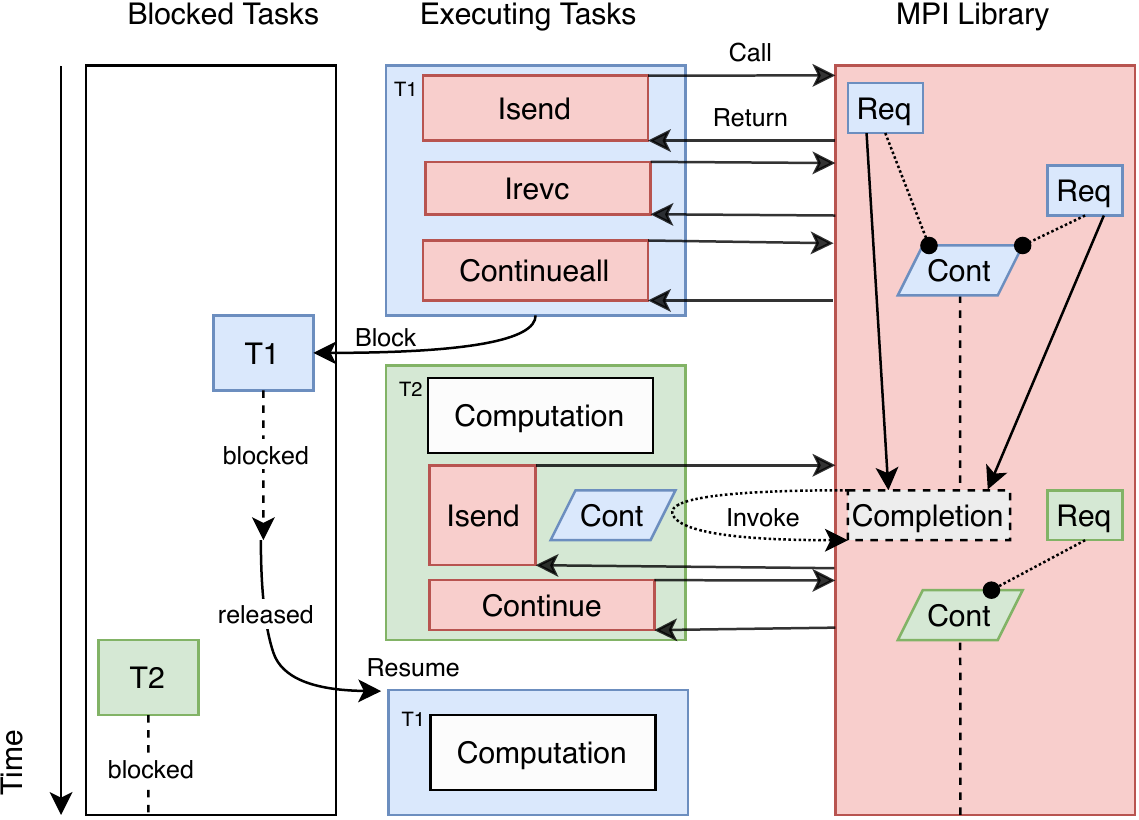}
\caption{The flow of control between tasks and MPI.}
\label{fig:control_flow}
\end{figure}

An example of a potential control flow between tasks and the MPI implementation is sketched in \autoref{fig:control_flow}.
Task \code{T1} (blue) is calling \code{MPI\_Isend} and \code{MPI\_Irecv} before attaching a continuation to the resulting requests.
The task is then blocked by means provided by the tasking library, allowing a second task \code{T2} (green) to be scheduled for execution.
After some computation, \code{T2} calls \code{MPI\_Isend}, in which the MPI library determines that both previously created requests have completed.
Consequently, the continuation is invoked in the context of \code{T2} from within the call to \code{MPI\_Isend}, which releases the blocked task \code{T1}.
After returning from the MPI call, \code{T2} attaches a continuation to its resulting request and blocks, allowing the execution of the just released task \code{T1} to resume.

\subsection{Usage in a Simple Example}
\label{sec:simple_example}

The example provided in \autoref{lst:mpi_continue_ex} uses MPI Continuations to release detached tasks in an OmpSs-2~\cite{BSC:2019:OmpSs2} application once communication operations have completed.
Looping over all fields of a local domain in each timestep, one task is created per field, which carries an output dependency on its \code{field} object (Line~\ref{line:task1}).
Inside each task, the received boundary data from the previous timestep is incorporated and a solver is applied to the field (potentially with nested tasks) before the local boundary is packed for sending (Lines~\ref{line:task3}--\ref{line:task4}).

Both the send and receive operations are initiated and a continuation is attached using \code{MPIX\_Continueall} (Lines~\ref{line:continue_reg1}--\ref{line:continue_reg2}).
OmpSs-2 offers direct access to API functions of the underlying \code{nanos6} runtime.
In this case, the event counter for the current task is retrieved, increased, and passed as the context of the continuation.
If all operation completed immediately, the event counter is decremented again.\footnote{The event counter has to be incremented first as the OmpSs-2 specification mandates that ``the user is responsible for not fulfilling events that the target task has still not bound.''~\cite[\S4.4]{BSC:2019:OmpSs2}}
Otherwise, the task will run to completion but its dependencies will not be released before the event counter is decremented in the continuation callback \code{continue\_cb} (Lines~\ref{line:continue_cb1}--\ref{line:continue_cb2}).

\begin{listing}
\begin{lstlisting}[numbers=left, breaklines=true, postbreak=\mbox{\textcolor{red}{$\hookrightarrow$}\space}]
int poll_mpi(MPI_Request *cont_req) { |\label{line:poll_mpi1}|
  int flag; /* result stored in flag not relevant here */
  MPI_Test(&flag, cont_req, MPI_STATUS_IGNORE);
  return false; /* signal that we should be called again */
}    |\label{line:poll_mpi2}|
void continue_cb(MPI_Status *status, void *task) { |\label{line:continue_cb1}|
  /* release the task's dependencies */
  nanos6_decrease_task_event_counter(task, 1);
} |\label{line:continue_cb2}|
void solve(int NT, int num_fields, field_t *fields[num_fields]) {
  /* create continuation request */
  MPI_Request cont_req;
  MPIX_Continue_init(&cont_req);
  /* register polling service */
  nanos6_register_polling_service("MPI", &poll_mpi, &cont_req); |\label{line:poll_mpi3}|

  for (int timestep = 0; timestep < NT; timestep++) {
    for (int i = 0; i < num_fields; ++i) {
      #pragma oss task depend(out: fields[i])   |\label{line:task1}|
      {
        /* unpack recv buffer, compute and pack send buffer */
        integrate_halo(fields[i]);        |\label{line:task3}|
        solve(fields[i]);
        save_boundary(fields[i]);            |\label{line:task4}|
        
        /* start send and recv */
        MPI_Request reqs[2];     |\label{line:continue_reg1}|
        MPI_Isend(fields[i]->sendbuf, ..., &reqs[0]);
        MPI_Irecv(fields[i]->recvbuf, ..., &reqs[1]);
        
        /* detach task if requests are active */
        void* task = nanos6_get_current_event_counter();
        nanos6_increase_current_task_event_counter(task, 1);
        /* attach continuation */
        int flag;
        MPIX_Continueall(2, reqs, &flag, &continue_cb, task, MPI_STATUSES_IGNORE, cont_req);
        if (flag) nanos6_decrease_task_event_counter(task, 1);  |\label{line:continue_reg2}\label{line:task2}|
  } } }
  /* wait for all tasks to complete and tear down */
  #pragma oss taskwait
  nanos6_unregister_polling_service("MPI", &poll_mpi, NULL);
  MPI_Request_free(&cont_req);
}
\end{lstlisting}
\caption{Simplified example using MPI Continuations in an iterative solver. Tasks are created for each field per timestep. Communication is initiated and tasks are detached. As soon as the communication completes, the task's dependencies are released and the field's next iteration may be scheduled.}
\label{lst:mpi_continue_ex}
\end{listing}

In order to ensure that all continuations are eventually completed, a \emph{polling service} is registered with the OmpSs-2 runtime (Line~\ref{line:poll_mpi3}).
This polling service is regularly invoked by the runtime and---in the function \code{mpi\_poll\_service} in Lines~\ref{line:poll_mpi1}--\ref{line:poll_mpi2}---calls \code{MPI\_Test} on the continuation request.\footnote{Note that OpenMP does not currently provide such an interface, which requires the user to create a \emph{progress task} or spawn a \emph{progress thread} that yield after testing.}
This will invoke all available continuations and eventually cause all tasks of the next timestep to become available for execution once the respective send and receive operations of the previous timestep have completed.

With only approximately 15 lines of code (including setup and tear-down), it is possible to integrate MPI communication with a task-parallel application to fully overlap communication and computation.
While this is, naturally, more code than would be required with TAMPI or fiber support in MPI, it is fully portable and allows an application to manage application-level concerns while leaving MPI-level concerns such as request management to MPI.
By making the interaction between non-blocking MPI operations and the task scheduler explicit, this approach ensures that both MPI and the task programming system support all required operations.

\subsection{Rationale}
\label{sec:rationale}
A scheme similar to the one proposed here could be implemented outside of the MPI library with an interface similar to TAMPI.
However, a main advantage of the integration with MPI is that the continuations can be invoked as soon as \emph{any} thread calls into MPI and determines the associated operations to be complete.
In more complex applications, this allows for the invocation of continuations set up by one part of the application during an MPI call issued in another part of the application, potentially reducing the time to release of blocked tasks and thus preventing thread starvation.

\subsubsection{Restrictions}
\label{sec:restrictions}
As stated earlier, continuations may be executed by application threads while executing communication functions in MPI.
Exceptions are \code{MPIX\_Continue[all]}, which may require the application to hold a mutex that it would then attempt to acquire again inside the continuation (see \autoref{sec:eval_latency} for an example).
While not prohibited, the use of blocking operations inside continuations should be avoided as it may cause a call to a nonblocking MPI procedure to block on an unrelated operation.
No other continuation may be invoked in MPI calls made within a continuation.

Continuations may not be invoked from within signal handlers inside the MPI implementation as that would restrict the application to using only async-signal-safe operations.
MPI implementation relying on signals to interact with the network hardware should defer the execution to the next invocation of an MPI procedure by any of the application threads or hand over to a progress thread.

\subsubsection{State Transitions}
\begin{figure}
\includegraphics[width=.9\columnwidth]{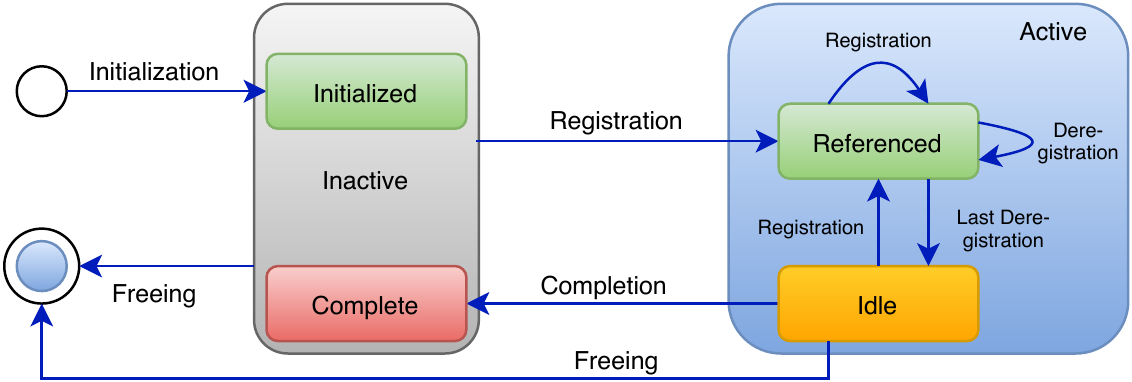}
\caption{State diagram of continuation requests.}
\label{fig:cont_state}
\end{figure}

Unlike existing request types in MPI, continuation requests (CR) do not represent a single operation but a set of operations.
Extending the semantics of persistent requests~\cite{Bangalore:2019:ECE}, \autoref{fig:cont_state} depicts the state diagram of CRs.
That state may change with every new registration or completion of a continuation.
A newly \emph{Initialized} or otherwise \emph{Inactive} CR becomes \emph{Active Referenced} when a continuation is registered.
It remains \emph{Active Referenced} if further continuations are registered.
Upon completion of continuations, they are deregistered from the CR.
The CR becomes \emph{Active Idle} upon the deregistration of the last active continuation.
An \emph{Active Idle} CR may become either \emph{Active Referenced} again if a new continuation is registered or \emph{Complete} if a \emph{Completion} function such as \code{MPI\_Test} is called on it.
It is possible to call \code{MPI\_Request\_free} on an active CR, in which case the CR cannot be used to register additional continuations and will be released as soon as all previously registered continuations have completed.

A continuation may be attached to a CR itself and registered with a separate CR, allowing applications to build complex patterns of notifications.
The new continuation will then be executed once all continuations registered with the first CR have completed.

\subsubsection{Thread-safety Requirements}

Multiple threads may safely register continuations with the same continuation request in parallel but only one thread may test or wait for its completion at any given time.
This allows for continuations to be registered concurrently without requiring mutual exclusion at the application level.

\subsubsection{Progress Requirements}
\label{sec:rationale_progress}

The availability of continuations is only one piece of the puzzle to reliably and portably integrate MPI with task-based applications.
The issue of progress in MPI is long-standing and while some implementations provide internal mechanisms for progressing communication (e.g., progress threads) this is not behavior users can rely on in all circumstances.
Thus, applications are still required to call into MPI to progress outstanding communication operations and---with the help of the proposed interface---invoke available continuation callbacks.
This proposal does not change the status quo in that regard.

It is thus left to the application layer to make sure that communication is progressing.
This can be achieved by regularly testing the continuation request, which could be done either through a recurring task, a polling service registered with the task runtime system (as shown for OmpSs-2 in \autoref{lst:mpi_continue_ex}), or by relying on a progress thread inside MPI.
The interface presented here is flexible enough to allow for its use in a wide range of circumstances and it defers to the application layer to choose the right strategy based on the available capabilities.

\section{Evaluation}
\label{sec:evaluation}
We will evaluate the proposed interface using the OSU latency benchmark.
We have ported its multi-threaded benchmark to use Argobots instead of pthreads.
We will then discuss a port of the NPB BT-MZ benchmark to \CC to compare TAMPI with MPI continuations used with both OmpSs-2 and OpenMP detached tasks.

All measurements were conducted on Vulcan, a small production NEC Linux cluster employing dual-socket 12-core Intel Xeon E5-2680v3 (Haswell) nodes connected using Mellanox ConnectX-3 cards.\footnote{Due to delays caused by the prevailing global COVID-19 situation, we are unable to report numbers on a larger system currently being commissioned at HLRS. However, early benchmarks indicated similar results to the ones presented here.}
The configurations of the used software are listed in \autoref{tab:software}.

All data points represent the mean of at least five repetitions, with the standard deviation plotted as error bars in the graph.

\begin{table}
\renewcommand{\arraystretch}{1.1}
\caption{Software configuration.}
\label{tab:software}
\begin{tabular}{llp{.55\columnwidth}}
\toprule
\textbf{Software} & \textbf{Version} & \textbf{Configuration/Remarks} \\
\midrule
\ompi & \code{git-0dc2325} & \code{--with-ucx=...} \\
MPICH & \code{git-256dea} & \code{--with-device=ch4:ucx}  \\
UCX & 1.8.0 & \code{--enable-mt} \\
GCC & 7.3.0 & \textit{site installation} \\
OmpSs-2 & 2019.11.2 & \code{--enable-ompss-2}, incl. patches mitigating task release performance issue \\
Argobots &  1.0rc2 & \textit{none} \\
Clang & \code{git-72edb79} & incl. fix \code{D79702} for detached tasks \\
PAPI  & 5.7.0 & \textit{site installation} \\
\bottomrule
\end{tabular}
\end{table}

\subsection{Implementation}
\begin{figure}
\centering
\includegraphics[width=.9\columnwidth, page=3]{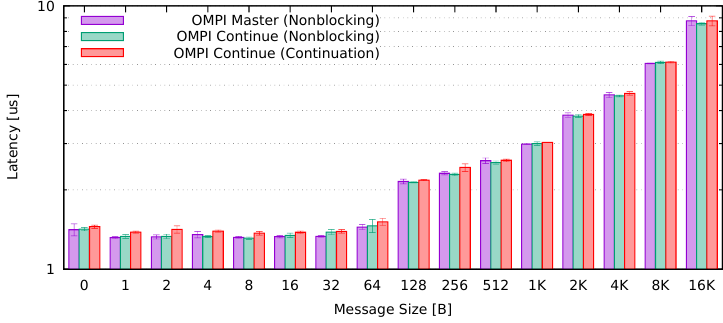}
\caption{Non-blocking P2P latency in vanilla \ompi, \ompi with continuations implemented but not used, and with continuations used to handle reply.}
\label{fig:osu_latency}
\end{figure}

We have implemented a proof-of-concept (PoC) of continuations within \ompi.\footnote{The PoC implementation is available at \url{https://github.com/devreal/ompi/tree/mpi-continue-master} (last accessed May 10, 2020).}
The fast-path for regular requests without a registered continuation adds 12 instructions over vanilla \ompi ($+2\%$ on a zero-byte message sent and received by the same process measured using PAPI).
Registration and invocation of an empty continuation requires 300 instructions in this scenario.
This is reflected in \autoref{fig:osu_latency}, which shows the non-blocking point-to-point latency measured for two processes on two different nodes, comparing the \ompi master branch against the PoC using traditional non-blocking send/recv followed by a wait as well as the use of continuations to send or receive the reply.
There is virtually no difference between vanilla \ompi and the PoC for non-blocking communication and a slight uptick in latency when using continuations (about 4\% for single-byte messages).
The difference vanishes for larger messages.

\subsection{OSU Latency with Argobots}
\label{sec:eval_latency}

In order to directly compare the continuations interface with the Argobots integration, we ported the multi-threaded OSU latency P2P benchmark (\code{osu\_latency\_mt}) to use Argobots instead of POSIX threads with blocking communication.\footnote{The implementation of the benchmarks can be found at \url{https://github.com/devreal/osu-abt-benchmarks} (Last accessed May 20, 2020).}
The code used to block and unblock Argobot fibers when using Continuations with non-blocking communication is listed in \autoref{lst:block_unblock_abt}.

The current Argobots integration in \ompi does not employ \ompi's synchronization mechanisms and instead relies on yielding inside the progress engine.
We thus had to configure the UCX PML to trigger the global progress engine unconditionally by passing \code{--mca pml\_ucx\_progress\_iterations 1} to \code{mpirun} as otherwise the latencies would have been an order of magnitude higher.
We have achieved similar latencies with a test-yield cycle inside the benchmark instead of using blocking communication.

\begin{listing}
\begin{lstlisting}[breaklines=true, postbreak=\mbox{\textcolor{red}{$\hookrightarrow$}\space}]
void block_fiber(fiber_state_t *fs, MPI_Request *req) {
    int flag;
    ABT_mutex_lock(fs->mtx);
    MPIX_Continue(&req, &flag, &unblock_fiber, fs, MPI_STATUS_IGNORE, cont_req);
    if (!flag) ABT_cond_wait(fs->cond, fs->mtx);
    ABT_mutex_unlock(fs->mtx);
}
int unblock_fiber(MPI_Status *status, void *data) {
    fiber_state_t *ts = (fiber_state_t *)data;
    ABT_mutex_lock(fs->mtx);
    ABT_cond_signal(fs->cond);
    ABT_mutex_unlock(fs->mtx);
    return MPI_SUCCESS;
}
\end{lstlisting}
\caption{%
Code to block and unblock an Argobot fiber.
The \code{fiber\_state\_t} structure contains a fiber-specific mutex and conditional variable.
Locking the mutex in \code{unblock\_fiber} is necessary to prevent signals from getting lost.%
}
\label{lst:block_unblock_abt}
\end{listing}

\begin{figure}
\begin{subfigure}{.48\columnwidth}
\centering
\includegraphics[width=\textwidth, page=11]{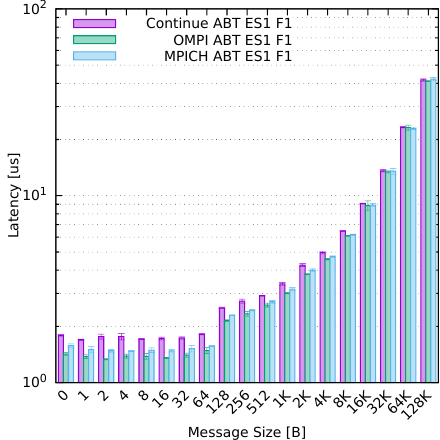}
\subcaption{1 thread, 1 fiber}
\label{fig:osu_latency_abt:e1t1}
\end{subfigure}
\hfill
\begin{subfigure}{.48\columnwidth}
\centering
\includegraphics[width=\textwidth, page=12]{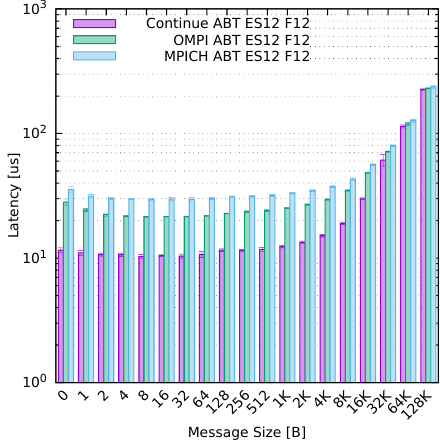}
\subcaption{12 threads, 12 fibers}
\label{fig:osu_latency_abt:e12t12}
\end{subfigure}

\begin{subfigure}{.48\columnwidth}
\centering
\includegraphics[width=\textwidth, page=13]{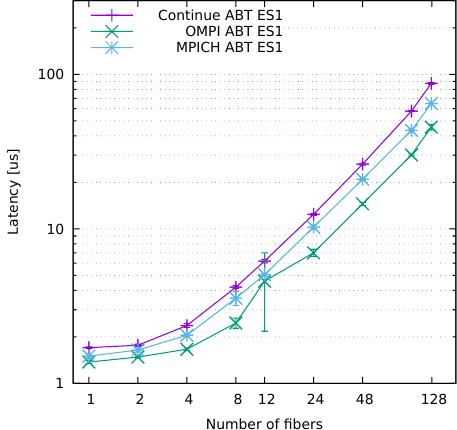}
\subcaption{Fiber scaling (1 thread, single-byte message)}
\label{fig:osu_latency_abt:e1}
\end{subfigure}
\hfill
\begin{subfigure}{.48\columnwidth}
\centering
\includegraphics[width=\textwidth, page=15]{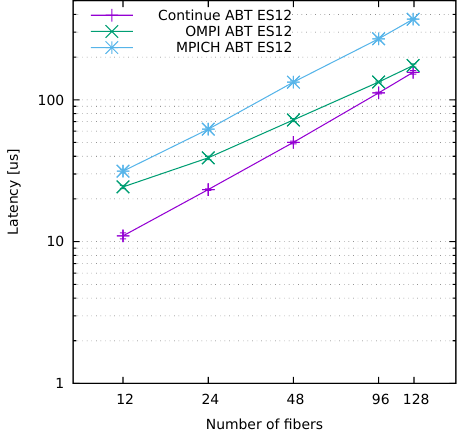}
\subcaption{Fiber scaling (12 threads, single-byte message)}
\label{fig:osu_latency_abt:e12}
\end{subfigure}

\caption{Argobots port of the \code{osu\_latency\_mt} P2P benchmark with different numbers of threads and fibers.}
\label{fig:osu_latency_abt}
\end{figure}

The results are depicted in \autoref{fig:osu_latency_abt} with different combinations of worker threads (\code{execution streams}) and fibers.
With a single thread and a single fiber (\autoref{fig:osu_latency_abt:e1t1}), the overhead of using conditional variables induces higher latencies than the yielding in \ompi's Argobots integration, leading to 23\% higher latencies for small messages.
This effect vanishes for larger messages starting at 32\,KB.
However, with 12 worker threads and one fiber per thread (\autoref{fig:osu_latency_abt:e12t12}), the use of conditional variables with continuations yields more than $2\times$ lower latency than the yielding in \ompi and $3\times$ lower latency than MPICH for small messages.

Scaling the number of fibers shown in \autoref{fig:osu_latency_abt:e1} and \autoref{fig:osu_latency_abt:e12} follows a similar trend: while for a single worker-thread using yield seems to be more efficient, the use of conditional variables is more efficient with 12 worker threads.
In contrast to the MPI implementation, the application has full knowledge of its configuration and can choose between yielding and blocking fibers.

\subsection{NPB BT-MZ}

In this section, we demonstrate the use of MPI Continuations in the context of the NAS Parallel Benchmark application BT-MZ, a multi-zone block tri-diagonal solver for the unsteady, compressible Navier Stokes equations on a three-dimensional mesh with two-dimensional domain decomposition~\cite{Wijngaart:2003:NPBMZ}.
Zones of different sizes are distributed across processes based on a static load-balancing scheme.
In each timestep and for each local zone, five computational steps are involved: forming the right-hand side, solving block-diagonal systems in each dimension \code{x}, \code{y}, and \code{z}, and updating the solution.
We use two representative problem sizes---classes C and D---for which the input configurations are listed in \autoref{tab:npb_problem_sizes}.
The reference implementation in Fortran uses OpenMP work-sharing constructs to parallelize nested loops during the updates to all local zones before collecting and exchanging all boundary data at the end of each timestep.
OpenMP parallelization is done over the outermost loop, which in most cases is the smallest dimension \code{z}, with the notable exception of the solution in the \code{z} dimension itself.

We have ported the Fortran version to \CC and implemented two variations using task-based concurrency.\footnote{The different variants of NPB BT-MZ discussed here are available at \url{https://github.com/devreal/npb3.3} (Last accessed May 20, 2020).}
The first variation uses tasks to overlap the computation of zones within a timestep in a fork-join approach in between boundary updates, replacing OpenMP parallel loops with loops generating tasks, with their execution coordinated using dependencies.
The second variant extends this to also perform the boundary exchange inside tasks, including the necessary MPI communication, effectively minimizing the coupling between zones to the contact point dependencies.

The second task-based variant requires some support from the tasking library to properly handle communicating tasks.
We use TAMPI in combination with OmpSs-2 as well as detached tasks in OpenMP available from Clang in its current trunk.\footnote{We attempted to use BOLT but in doing so uncovered a design flaw that stems from the current thread/fiber ambiguity and invalidates both the use of Argobots conditional variables as depicted in \autoref{lst:block_unblock_abt} and any Argobots MPI integration. An analysis can be found at \url{https://github.com/pmodels/bolt/issues/51} (last accessed May 14, 2020). 
}
We have attempted to use \code{taskyield} within OpenMP tasks using the Clang implementation, which resulted in deadlocks due to the restricted semantics of task-yield in OpenMP~\cite{Schuchart:2018:TIT}.
The implementation using detached tasks spawns a progress thread that tests the continuation request before calling \code{usleep} to yield the processor.

\begin{table}
\caption{Problem sizes in the NPB BT-MZ benchmark~\cite{NPB:2019:ProbSizes}.}
\label{tab:npb_problem_sizes}
\centering
\begin{tabular}{cm{.15\columnwidth}cm{.22\columnwidth}m{.15\columnwidth}}
\toprule
\textbf{Class} & \textbf{Zones} & \textbf{Iterations} & \raggedright \centering \textbf{Grid Size} &  \centering\arraybackslash \textbf{Memory} \\
\midrule
C  & $16 \times 16$ & 200 & \centering$480 \times 320 \times 28$ & \raggedleft\arraybackslash 0.8\,GB \\
D  & $32 \times 32$ & 250 & \centering$1632 \times 1216 \times 34$ & \raggedleft\arraybackslash 12.8\,GB \\
\bottomrule
\end{tabular}
\end{table}

Given the limited concurrency in the OpenMP loops, we present the \emph{minimum} of the average runtime of 5 repetitions achieved using 1, 2, or 4 processes per nodes (24, 12, and 6 threads, respectively).
All codes have been compiled with the flags \code{-O2 -march=haswell -mcmodel=medium}.
The \CC variants were compiled using the Clang compiler while the Fortran variant was compiled using GCC's \code{gfortran}.
OpenMP thread affinity was enabled (\code{OMP\_PROC\_BIND=true}).
Process binding has been achieved using \ompi's \code{--map-by node:PE=\${nthreads} --bind-to core} arguments to \code{mpirun}.

\begin{figure}
\begin{subfigure}{.48\columnwidth}
\centering
\includegraphics[width=\textwidth, page=5]{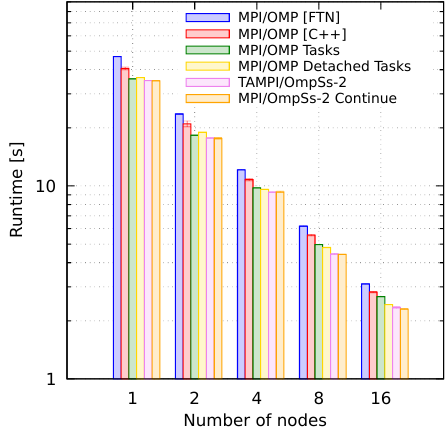}
\subcaption{Class C Runtime}
\label{fig:npb_bt_mz:runtime:c}
\end{subfigure}
\hfill
\begin{subfigure}{.48\columnwidth}
\centering
\includegraphics[width=\textwidth, page=6]{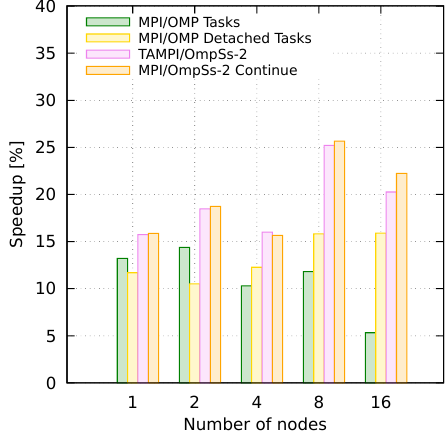}
\subcaption{Class C Speedup}
\label{fig:npb_bt_mz:speedup:c}
\end{subfigure}

\begin{subfigure}{.48\columnwidth}
\centering
\includegraphics[width=\textwidth, page=5]{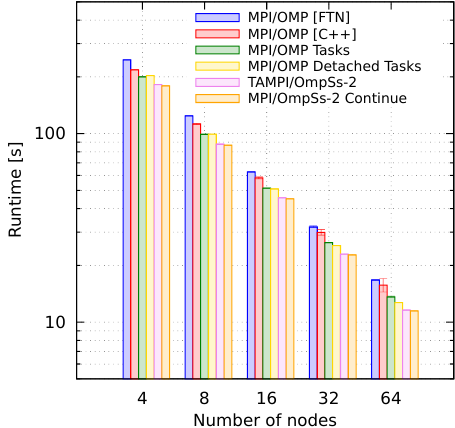}
\subcaption{Class D Runtime}
\label{fig:npb_bt_mz:runtime:d}
\end{subfigure}
\hfill
\begin{subfigure}{.48\columnwidth}
\centering
\includegraphics[width=\textwidth, page=6]{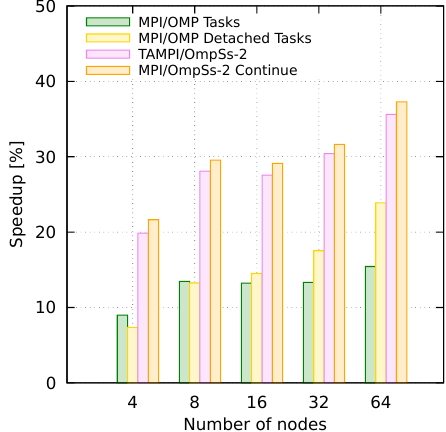}
\subcaption{Class D Speedup}
\label{fig:npb_bt_mz:speedup:d}
\end{subfigure}

\caption{NPB BT-MZ benchmark results. Speedups are relative to the \CC port.}
\label{fig:npb_bt_mz}
\end{figure}

\autoref{fig:npb_bt_mz} provides results for classes C and D.
Most notably, the task-based variants with \emph{fine-grained} MPI communication (OpenMP detached tasks, OmpSs-2) outperform the variants using OpenMP loop-based worksharing and bulk communication by at least 15\% (C) or 24\% (D) at scale.
It is also notable that the OmpSs-2 variants yield by far the highest performance, even though their task-structure is similar to that using OpenMP detached tasks.
We attribute this to potential degradation of parallelism in Clang's \code{libomp} if tasks waiting in \code{taskwait} constructs (required to wait for nested tasks) are stacked up on the same thread, similar to issues reported for the \code{taskyield} directive~\cite{Schuchart:2018:TIT}.\footnote{We believe that the use of \fncode{untied} tasks may improve the situation by avoiding the stacking but a long-standing issue in Clang's \fncode{libomp} prevents their use: \url{https://bugs.llvm.org/show_bug.cgi?id=37671} (Last accessed May 19, 2020).}
The usage of MPI Continuations with OmpSs-2 provides a slight speedup of about 1.5\% over TAMPI for class D, suggesting a more efficient communication management.

\section{Related Work}

Multiple efforts have been made in the past to improve specific aspects of the interaction between multi-threaded programming models and MPI~\cite{Hoefler:2010:EMS,Dinan:2013:EMI,Grant:2019:FPM}. 

Extended generalized requests have been proposed to encapsulate operations within requests and allow the MPI implementation to progress them through a callback~\cite{Latham:2007:EMG}.
Similar to the interface proposed here, extended generalized requests provide \emph{inversion of control} (IoC)~\cite{Mattsson:96:OOF}: application-level code is called by the MPI library in inversion of the usual call relation.
However, these requests are still polling-based and cannot be used for completion notification.

IoC is also used in the upcoming MPI\_T events interface, allowing applications and tools to be notified through callbacks about events from within the MPI library~\cite{Hermanns:2019:MPIT}.
However, the availability of events is implementation-specific and callbacks can only be registered for \emph{event types}, as opposed to specific operations.
The MPI\_T interface has been used to structure the interaction between MPI and task-based programming models, still requiring the mapping between MPI requests and task-information in the application layer~\cite{Castillo:2019:OCC}.

Continuations are not a new concept in the context of low-level communication libraries: UCX allows passing callback notification functions when initiating a non-blocking operation~\cite{UCX:1.6}.
In contrast, other interfaces such as libfabric~\cite{OFI:2017}, Portals~\cite{Barrett:2018:Portals4}, and uGNI~\cite{Pritchard:2011:uGNI} rely on completion queues from which notifications are popped. 
While this would be a viable concept in MPI as well, we believe that continuations provide greater flexibility at the application level. 

Another alternative would be a system similar to \emph{scheduler activations} in OS kernels~\cite{Anderson:1991:SAE}, where MPI would signal to some application-level entity that the current thread or fiber would block inside a call.
This has been proposed for the OpenShmem standard~\cite{Rahman:SCO:2020}.
However, its use is limited to tasking models that have strong rescheduling guarantees and would not work with OpenMP detached tasks.

\emph{Hardware-triggered operations} are being used to react to events in the network with low latency and to offload computation to the network interface card~\cite{Hoefler:2017:SHS,Schonbein:2019:INCA,Islam:2019:MUH}.
However, in-network computing is focused on operations on the data stream and likely cannot be utilized to execute arbitrary code in user-space, e.g., call into fiber libraries.
Exploring the boundaries of integrating these systems with the interface proposed here remains as future work.

The integration of fiber libraries into MPI has been proposed as a replacement for POSIX thread support to enable the use of blocking MPI calls inside tasks and fibers~\cite{Lu:2015:MUO}.
This has been discussed in \autoref{sec:aspects}.
Other proposals provide runtime-specific wrapper libraries around MPI, transforming blocking MPI calls to non-blocking calls and testing for completion before releasing a fiber or task~\cite{Stark:2014:EECS,Sala:2018:IIM,Sala:2019:TAMPI}.
These proposals fall short of providing a portable interface that can be used with arbitrary asynchronous programming models.

A proposal for completion callbacks for MPI request objects has been discussed in the context of the MPI Forum over a decade ago and rejected due to its broad focus~\cite{Hoefler:2008:RCCF}.
In contrast, the proposal outlined here avoids some of the pitfalls such as callback contention, ownership of requests, and unclear progress semantics.

The interface proposed here is especially well-suited for node-local asynchronous programming models such as OpenMP~\cite{openmp5.0}, OmpSs-2~\cite{BSC:2019:OmpSs2}, and Intel TBB~\cite{Reinders:2007:TBB} or in combination with cooperatively scheduled fiber libraries such as Argobots~\cite{Seo:2018:AAL} or Qthreads~\cite{Wheeler:2008:QAA} to ease the communication management burden users are facing.
However, it may also be useful in distributed runtime systems such as PaRSEC~\cite{DAGuE:2011}, HPX~\cite{hpx:2014}, StarPU~\cite{Augonnet:2011}, and DASH~\cite{Schuchart:2019:GTD} to simplify internal communication handling.
An evaluation of the integration with runtime systems remains as future work.

\section{Conclusions}
\label{sec:conclusions}

This paper has made the argument that the integration of fiber libraries into MPI implementations for the sake of application programmability is neither warranted by the MPI standard nor beneficial in terms of maintaining an ecosystem of portable applications.
In order to tackle the challenges of asynchronous activities communicating through MPI, we have proposed an extension to the MPI standard that allows users to register continuations with active operations, which will be invoked once the MPI library determines the operation to be complete.
We have shown that this interface is easy to use and flexible enough to accommodate three different tasking and fiber models (OpenMP detached tasks, OmpSs-2, and Argobots).
The overhead of our proof-of-concept implementation is sufficiently low to not impact existing applications and its use shows similar or improved performance over existing approaches.

The interface has been designed with higher-level abstractions such as \CC futures for MPI in mind but it remains as future work to provide an actual implementation on top of MPI Continuations.
It is also an open point of discussion whether it is desirable to allow users to limit the context in which certain continuations may be invoked to mitigate the impact of expensive continuations on latency-sensitive parts of the application.

\begin{acks}
This research was in parts funded by the \grantsponsor{DFG}{German Research Foundation (DFG)}{http://www.dfg.org} through the \grantnum{DFG:1648}{German Priority Programme 1648 Software for Exascale Computing (SPPEXA)} in the SmartDASH project and through financial support by the Federal Ministry of Education and Research, Germany, grant number 01IH16008B (project TaLPas).

The authors would like to thank Vicenç Beltran Querol and Kevin Sala at Barcelona Supercomputing Center for their support with OmpSs-2 and Joachim Protze for his support with Clang's \code{libomp}.
\end{acks}

\bibliographystyle{ACM-Reference-Format}
\bibliography{references}


\begin{thebibliography}{61}


\ifx \showCODEN    \undefined \def \showCODEN     #1{\unskip}     \fi
\ifx \showDOI      \undefined \def \showDOI       #1{#1}\fi
\ifx \showISBNx    \undefined \def \showISBNx     #1{\unskip}     \fi
\ifx \showISBNxiii \undefined \def \showISBNxiii  #1{\unskip}     \fi
\ifx \showISSN     \undefined \def \showISSN      #1{\unskip}     \fi
\ifx \showLCCN     \undefined \def \showLCCN      #1{\unskip}     \fi
\ifx \shownote     \undefined \def \shownote      #1{#1}          \fi
\ifx \showarticletitle \undefined \def \showarticletitle #1{#1}   \fi
\ifx \showURL      \undefined \def \showURL       {\relax}        \fi
\providecommand\bibfield[2]{#2}
\providecommand\bibinfo[2]{#2}
\providecommand\natexlab[1]{#1}
\providecommand\showeprint[2][]{arXiv:#2}

\bibitem[\protect\citeauthoryear{Adya, Howell, Theimer, Bolosky, and
  Douceur}{Adya et~al\mbox{.}}{2002}]%
        {Adya:2002:CTM}
\bibfield{author}{\bibinfo{person}{Atul Adya}, \bibinfo{person}{Jon Howell},
  \bibinfo{person}{Marvin Theimer}, \bibinfo{person}{William~J. Bolosky}, {and}
  \bibinfo{person}{John~R. Douceur}.} \bibinfo{year}{2002}\natexlab{}.
\newblock \showarticletitle{Cooperative Task Management Without Manual Stack
  Management}. In \bibinfo{booktitle}{\emph{Proceedings of the General Track of
  the Annual Conference on USENIX Annual Technical Conference}}
  \emph{(\bibinfo{series}{ATEC ’02})}. \bibinfo{publisher}{USENIX
  Association}.
\newblock
\showISBNx{1880446006}
\urldef\tempurl%
\url{https://doi.org/10.5555/647057.713851}
\showDOI{\tempurl}


\bibitem[\protect\citeauthoryear{Anderson, Bershad, Lazowska, and
  Levy}{Anderson et~al\mbox{.}}{1991}]%
        {Anderson:1991:SAE}
\bibfield{author}{\bibinfo{person}{Thomas~E. Anderson},
  \bibinfo{person}{Brian~N. Bershad}, \bibinfo{person}{Edward~D. Lazowska},
  {and} \bibinfo{person}{Henry~M. Levy}.} \bibinfo{year}{1991}\natexlab{}.
\newblock \showarticletitle{Scheduler Activations: Effective Kernel Support for
  the User-Level Management of Parallelism}. In
  \bibinfo{booktitle}{\emph{Proceedings of the Thirteenth ACM Symposium on
  Operating Systems Principles}} \emph{(\bibinfo{series}{SOSP ’91})}.
  \bibinfo{publisher}{Association for Computing Machinery}.
\newblock
\urldef\tempurl%
\url{https://doi.org/10.1145/121132.121151}
\showDOI{\tempurl}


\bibitem[\protect\citeauthoryear{Augonnet, Thibault, Namyst, and
  Wacrenier}{Augonnet et~al\mbox{.}}{2011}]%
        {Augonnet:2011}
\bibfield{author}{\bibinfo{person}{C\'{e}dric Augonnet},
  \bibinfo{person}{Samuel Thibault}, \bibinfo{person}{Raymond Namyst}, {and}
  \bibinfo{person}{Pierre-Andr\'{e} Wacrenier}.}
  \bibinfo{year}{2011}\natexlab{}.
\newblock \showarticletitle{{StarPU: A Unified Platform for Task Scheduling on
  Heterogeneous Multicore Architectures}}.
\newblock \bibinfo{journal}{\emph{Concurrent Computing: Practice and
  Experience}} \bibinfo{volume}{23}, \bibinfo{number}{2} (\bibinfo{date}{Feb.}
  \bibinfo{year}{2011}), \bibinfo{pages}{187--198}.
\newblock
\urldef\tempurl%
\url{https://doi.org/10.1002/cpe.1631}
\showDOI{\tempurl}


\bibitem[\protect\citeauthoryear{Bangalore, Rabenseifner, Holmes, Jaeger,
  Mercier, Blaas-Schenner, and Skjellum}{Bangalore et~al\mbox{.}}{2019}]%
        {Bangalore:2019:ECE}
\bibfield{author}{\bibinfo{person}{Purushotham~V. Bangalore},
  \bibinfo{person}{Rolf Rabenseifner}, \bibinfo{person}{Daniel~J. Holmes},
  \bibinfo{person}{Julien Jaeger}, \bibinfo{person}{Guillaume Mercier},
  \bibinfo{person}{Claudia Blaas-Schenner}, {and} \bibinfo{person}{Anthony
  Skjellum}.} \bibinfo{year}{2019}\natexlab{}.
\newblock \showarticletitle{Exposition, Clarification, and Expansion of MPI
  Semantic Terms and Conventions: Is a Nonblocking MPI Function Permitted to
  Block?}. In \bibinfo{booktitle}{\emph{Proceedings of the 26th European MPI
  Users’ Group Meeting}} (Z\"{u}rich, Switzerland)
  \emph{(\bibinfo{series}{EuroMPI ’19})}. \bibinfo{publisher}{Association for
  Computing Machinery}, \bibinfo{address}{New York, NY, USA}, Article
  \bibinfo{articleno}{2}, \bibinfo{numpages}{10}~pages.
\newblock
\showISBNx{9781450371759}
\urldef\tempurl%
\url{https://doi.org/10.1145/3343211.3343213}
\showDOI{\tempurl}


\bibitem[\protect\citeauthoryear{Barrett, Brightwell, Grant, Hemmert, Pedretti,
  Wheeler, Underwood, Riesen, Hoefler, Maccabe, and Hudson}{Barrett
  et~al\mbox{.}}{2018}]%
        {Barrett:2018:Portals4}
\bibfield{author}{\bibinfo{person}{Brian~W. Barrett}, \bibinfo{person}{Ron
  Brightwell}, \bibinfo{person}{Ryan~E. Grant}, \bibinfo{person}{Scott
  Hemmert}, \bibinfo{person}{Kevin Pedretti}, \bibinfo{person}{Kyle Wheeler},
  \bibinfo{person}{Keith Underwood}, \bibinfo{person}{Rolf Riesen},
  \bibinfo{person}{Torsten Hoefler}, \bibinfo{person}{Arthur~B. Maccabe}, {and}
  \bibinfo{person}{Trammell Hudson}.} \bibinfo{year}{2018}\natexlab{}.
\newblock \bibinfo{booktitle}{\emph{The Portals 4.2 Network Programming
  Interface}}.
\newblock \bibinfo{type}{{T}echnical {R}eport} SAND2018-12790.
  \bibinfo{institution}{Sandia National Laboratories}.
\newblock


\bibitem[\protect\citeauthoryear{Bernholdt, Boehm, Bosilca, Venkata, Grant,
  Naughton, Pritchard, Schulz, and Vallee}{Bernholdt et~al\mbox{.}}{2018}]%
        {Berholdt:2018:MPIUsage}
\bibfield{author}{\bibinfo{person}{David~E. Bernholdt}, \bibinfo{person}{Swen
  Boehm}, \bibinfo{person}{George Bosilca}, \bibinfo{person}{Manjunath~Grentla
  Venkata}, \bibinfo{person}{Ryan~E. Grant}, \bibinfo{person}{Thomas Naughton},
  \bibinfo{person}{Howard~P. Pritchard}, \bibinfo{person}{Martin Schulz}, {and}
  \bibinfo{person}{Geoffroy~R. Vallee}.} \bibinfo{year}{2018}\natexlab{}.
\newblock \showarticletitle{A Survey of MPI Usage in the US Exascale Computing
  Project}.
\newblock \bibinfo{journal}{\emph{Concurrency Computation: Practice and
  Experience}} (\bibinfo{date}{09-2018} \bibinfo{year}{2018}).
\newblock
\urldef\tempurl%
\url{https://doi.org/10.1002/cpe.4851}
\showDOI{\tempurl}


\bibitem[\protect\citeauthoryear{{Bosilca}, {Bouteiller}, {Danalis}, {Herault},
  {Lemarinier}, and {Dongarra}}{{Bosilca} et~al\mbox{.}}{2011}]%
        {DAGuE:2011}
\bibfield{author}{\bibinfo{person}{G. {Bosilca}}, \bibinfo{person}{A.
  {Bouteiller}}, \bibinfo{person}{A. {Danalis}}, \bibinfo{person}{T.
  {Herault}}, \bibinfo{person}{P. {Lemarinier}}, {and} \bibinfo{person}{J.
  {Dongarra}}.} \bibinfo{year}{2011}\natexlab{}.
\newblock \showarticletitle{DAGuE: A Generic Distributed DAG Engine for High
  Performance Computing}. In \bibinfo{booktitle}{\emph{2011 IEEE International
  Symposium on Parallel and Distributed Processing Workshops and Phd Forum}}.
  \bibinfo{pages}{1151--1158}.
\newblock
\urldef\tempurl%
\url{https://doi.org/10.1109/IPDPS.2011.281}
\showDOI{\tempurl}


\bibitem[\protect\citeauthoryear{Castillo, Jain, Casas, Moreto, Schulz,
  Beivide, Valero, and Bhatele}{Castillo et~al\mbox{.}}{2019}]%
        {Castillo:2019:OCC}
\bibfield{author}{\bibinfo{person}{Emilio Castillo}, \bibinfo{person}{Nikhil
  Jain}, \bibinfo{person}{Marc Casas}, \bibinfo{person}{Miquel Moreto},
  \bibinfo{person}{Martin Schulz}, \bibinfo{person}{Ramon Beivide},
  \bibinfo{person}{Mateo Valero}, {and} \bibinfo{person}{Abhinav Bhatele}.}
  \bibinfo{year}{2019}\natexlab{}.
\newblock \showarticletitle{Optimizing Computation-Communication Overlap in
  Asynchronous Task-Based Programs}. In \bibinfo{booktitle}{\emph{Proceedings
  of the ACM International Conference on Supercomputing}}
  \emph{(\bibinfo{series}{ICS ’19})}. \bibinfo{publisher}{Association for
  Computing Machinery}.
\newblock
\urldef\tempurl%
\url{https://doi.org/10.1145/3330345.3330379}
\showDOI{\tempurl}


\bibitem[\protect\citeauthoryear{Center}{Center}{2020}]%
        {BSC:2019:OmpSs2}
\bibfield{author}{\bibinfo{person}{Barcelona~Supercomputing Center}.}
  \bibinfo{year}{2020}\natexlab{}.
\newblock \bibinfo{booktitle}{\emph{OmpSs-2 Specification}}.
\newblock \bibinfo{type}{{T}echnical {R}eport}.
\newblock
\urldef\tempurl%
\url{https://pm.bsc.es/ftp/ompss-2/doc/spec/OmpSs-2-Specification.pdf}
\showURL{%
\tempurl}
\newblock
\shownote{Last accessed May 12, 2020.}


\bibitem[\protect\citeauthoryear{Dai\ss{}, Amini, Biddiscombe, Diehl, Frank,
  Huck, Kaiser, Marcello, Pfander, and Pf\"{u}ger}{Dai\ss{}
  et~al\mbox{.}}{2019}]%
        {Daliss:2019:FPD}
\bibfield{author}{\bibinfo{person}{Gregor Dai\ss{}}, \bibinfo{person}{Parsa
  Amini}, \bibinfo{person}{John Biddiscombe}, \bibinfo{person}{Patrick Diehl},
  \bibinfo{person}{Juhan Frank}, \bibinfo{person}{Kevin Huck},
  \bibinfo{person}{Hartmut Kaiser}, \bibinfo{person}{Dominic Marcello},
  \bibinfo{person}{David Pfander}, {and} \bibinfo{person}{Dirk Pf\"{u}ger}.}
  \bibinfo{year}{2019}\natexlab{}.
\newblock \showarticletitle{From Piz Daint to the Stars: Simulation of Stellar
  Mergers Using High-Level Abstractions}. In
  \bibinfo{booktitle}{\emph{Proceedings of the International Conference for
  High Performance Computing, Networking, Storage and Analysis}}
  \emph{(\bibinfo{series}{SC ’19})}.
\newblock
\urldef\tempurl%
\url{https://doi.org/10.1145/3295500.3356221}
\showDOI{\tempurl}


\bibitem[\protect\citeauthoryear{der Wijngaart and Jin}{der Wijngaart and
  Jin}{2003}]%
        {Wijngaart:2003:NPBMZ}
\bibfield{author}{\bibinfo{person}{Rob F.~Van der Wijngaart} {and}
  \bibinfo{person}{Haoqiang Jin}.} \bibinfo{year}{2003}\natexlab{}.
\newblock \bibinfo{booktitle}{\emph{{NAS Parallel Benchmarks, Multi-Zone
  Versions}}}.
\newblock \bibinfo{type}{{T}echnical {R}eport} NAS-03-010.
  \bibinfo{institution}{NASA Advanced Supercomputing (NAS) Division}.
\newblock


\bibitem[\protect\citeauthoryear{Dinan, Balaji, Goodell, Miller, Snir, and
  Thakur}{Dinan et~al\mbox{.}}{2013}]%
        {Dinan:2013:EMI}
\bibfield{author}{\bibinfo{person}{James Dinan}, \bibinfo{person}{Pavan
  Balaji}, \bibinfo{person}{David Goodell}, \bibinfo{person}{Douglas Miller},
  \bibinfo{person}{Marc Snir}, {and} \bibinfo{person}{Rajeev Thakur}.}
  \bibinfo{year}{2013}\natexlab{}.
\newblock \showarticletitle{Enabling MPI Interoperability through Flexible
  Communication Endpoints}. In \bibinfo{booktitle}{\emph{Proceedings of the
  20th European MPI Users’ Group Meeting}} \emph{(\bibinfo{series}{EuroMPI
  ’13})}. \bibinfo{publisher}{Association for Computing Machinery}.
\newblock
\urldef\tempurl%
\url{https://doi.org/10.1145/2488551.2488553}
\showDOI{\tempurl}


\bibitem[\protect\citeauthoryear{Duran, Eduard, Badia, Larbarta, Martinell,
  Martorell, and Plana}{Duran et~al\mbox{.}}{2011}]%
        {ompss2011}
\bibfield{author}{\bibinfo{person}{Alejandro Duran}, \bibinfo{person}{Ayguadem
  Eduard}, \bibinfo{person}{Rosa~M. Badia}, \bibinfo{person}{Jesus Larbarta},
  \bibinfo{person}{Luis Martinell}, \bibinfo{person}{Xavier Martorell}, {and}
  \bibinfo{person}{Judit Plana}.} \bibinfo{year}{2011}\natexlab{}.
\newblock \showarticletitle{{OmpSs: A proposal for programming heterogeneous
  multi-core architectures}}.
\newblock \bibinfo{journal}{\emph{Parallel Processing Letters}}
  \bibinfo{volume}{21}, \bibinfo{number}{02} (\bibinfo{year}{2011}),
  \bibinfo{pages}{173--193}.
\newblock
\urldef\tempurl%
\url{https://doi.org/10.1142/S0129626411000151}
\showDOI{\tempurl}


\bibitem[\protect\citeauthoryear{Engelschall}{Engelschall}{2006}]%
        {GNUPth}
\bibfield{author}{\bibinfo{person}{Ralf~S. Engelschall}.}
  \bibinfo{year}{2006}\natexlab{}.
\newblock \bibinfo{booktitle}{\emph{GNU Pth---The GNU Portable Threads}}.
\newblock \bibinfo{type}{{T}echnical {R}eport}.
\newblock
\urldef\tempurl%
\url{https://www.gnu.org/software/pth/pth-manual.html}
\showURL{%
\tempurl}
\newblock
\shownote{Last accessed April 24, 2020.}


\bibitem[\protect\citeauthoryear{Friedman, Haynes, and Kohlbecker}{Friedman
  et~al\mbox{.}}{1984}]%
        {Friedman:1984:PWC}
\bibfield{author}{\bibinfo{person}{Daniel~P. Friedman},
  \bibinfo{person}{Christopher~T. Haynes}, {and} \bibinfo{person}{Eugene
  Kohlbecker}.} \bibinfo{year}{1984}\natexlab{}.
\newblock \showarticletitle{Programming with Continuations}. In
  \bibinfo{booktitle}{\emph{Program Transformation and Programming
  Environments}}, \bibfield{editor}{\bibinfo{person}{Peter Pepper}} (Ed.).
  \bibinfo{publisher}{Springer Berlin Heidelberg}.
\newblock
\urldef\tempurl%
\url{https://doi.org/10.1007/978-3-642-46490-4_23}
\showDOI{\tempurl}


\bibitem[\protect\citeauthoryear{Goodspeed and Kowalke}{Goodspeed and
  Kowalke}{2014}]%
        {Goodspeed:2014:DCF}
\bibfield{author}{\bibinfo{person}{Nat Goodspeed} {and} \bibinfo{person}{Oliver
  Kowalke}.} \bibinfo{year}{2014}\natexlab{}.
\newblock \bibinfo{booktitle}{\emph{Distinguishing coroutines and fibers}}.
\newblock \bibinfo{type}{{T}echnical {R}eport} N4024.
\newblock
\urldef\tempurl%
\url{http://www.open-std.org/jtc1/sc22/wg21/docs/papers/2014/n4024.pdf}
\showURL{%
\tempurl}
\newblock
\shownote{Last accessed April 24, 2020.}


\bibitem[\protect\citeauthoryear{Grant, Dosanjh, Levenhagen, Brightwell, and
  Skjellum}{Grant et~al\mbox{.}}{2019}]%
        {Grant:2019:FPM}
\bibfield{author}{\bibinfo{person}{Ryan~E. Grant}, \bibinfo{person}{Matthew
  G.~F. Dosanjh}, \bibinfo{person}{Michael~J. Levenhagen}, \bibinfo{person}{Ron
  Brightwell}, {and} \bibinfo{person}{Anthony Skjellum}.}
  \bibinfo{year}{2019}\natexlab{}.
\newblock \showarticletitle{Finepoints: Partitioned Multithreaded MPI
  Communication}. In \bibinfo{booktitle}{\emph{High Performance Computing}},
  \bibfield{editor}{\bibinfo{person}{Mich{\`e}le Weiland},
  \bibinfo{person}{Guido Juckeland}, \bibinfo{person}{Carsten Trinitis}, {and}
  \bibinfo{person}{Ponnuswamy Sadayappan}} (Eds.). \bibinfo{publisher}{Springer
  International Publishing}, \bibinfo{address}{Cham},
  \bibinfo{pages}{330--350}.
\newblock
\showISBNx{978-3-030-20656-7}


\bibitem[\protect\citeauthoryear{Gustafsson, Laksberg, Sutter, and
  Mithani}{Gustafsson et~al\mbox{.}}{2014}]%
        {CPP:2014:stdfuture}
\bibfield{author}{\bibinfo{person}{Niklas Gustafsson}, \bibinfo{person}{Artur
  Laksberg}, \bibinfo{person}{Herb Sutter}, {and} \bibinfo{person}{Sana
  Mithani}.} \bibinfo{year}{2014}\natexlab{}.
\newblock \bibinfo{booktitle}{\emph{N3857: Improvements to std::future<T> and
  Related APIs}}.
\newblock \bibinfo{type}{{T}echnical {R}eport} N3857.
\newblock


\bibitem[\protect\citeauthoryear{Hermanns, Hjelm, Knobloch, Mohror, and
  Schulz}{Hermanns et~al\mbox{.}}{2019}]%
        {Hermanns:2019:MPIT}
\bibfield{author}{\bibinfo{person}{Marc-André Hermanns},
  \bibinfo{person}{Nathan~T. Hjelm}, \bibinfo{person}{Michael Knobloch},
  \bibinfo{person}{Kathryn Mohror}, {and} \bibinfo{person}{Martin Schulz}.}
  \bibinfo{year}{2019}\natexlab{}.
\newblock \showarticletitle{The MPI\_T events interface: An early evaluation
  and overview of the interface}.
\newblock \bibinfo{journal}{\emph{Parallel Comput.}} (\bibinfo{year}{2019}).
\newblock
\urldef\tempurl%
\url{https://doi.org/10.1016/j.parco.2018.12.006}
\showDOI{\tempurl}


\bibitem[\protect\citeauthoryear{Hjelm, Dosanjh, Grant, Groves, Bridges, and
  Arnold}{Hjelm et~al\mbox{.}}{2018}]%
        {Hjelm:2018:IMM}
\bibfield{author}{\bibinfo{person}{Nathan Hjelm}, \bibinfo{person}{Matthew
  G.~F. Dosanjh}, \bibinfo{person}{Ryan~E. Grant}, \bibinfo{person}{Taylor
  Groves}, \bibinfo{person}{Patrick Bridges}, {and} \bibinfo{person}{Dorian
  Arnold}.} \bibinfo{year}{2018}\natexlab{}.
\newblock \showarticletitle{{Improving MPI Multi-threaded RMA Communication
  Performance}}. In \bibinfo{booktitle}{\emph{Proceedings of the 47th
  International Conference on Parallel Processing}}
  \emph{(\bibinfo{series}{ICPP 2018})}. \bibinfo{publisher}{ACM},
  \bibinfo{pages}{58:1--58:11}.
\newblock
\urldef\tempurl%
\url{https://doi.org/10.1145/3225058.3225114}
\showDOI{\tempurl}


\bibitem[\protect\citeauthoryear{Hoefler}{Hoefler}{2008}]%
        {Hoefler:2008:RCCF}
\bibfield{author}{\bibinfo{person}{Torsten Hoefler}.}
  \bibinfo{year}{2008}\natexlab{}.
\newblock \bibinfo{title}{Request Completion Callback Function}.
\newblock \bibinfo{howpublished}{MPI Forum Discussion}.
\newblock
\newblock
\shownote{Archived at
  \url{https://github.com/mpi-forum/mpi-forum-historic/issues/26}, last
  accessed July 6, 2020.}


\bibitem[\protect\citeauthoryear{Hoefler, Bronevetsky, Barrett, de~Supinski,
  and Lumsdaine}{Hoefler et~al\mbox{.}}{2010}]%
        {Hoefler:2010:EMS}
\bibfield{author}{\bibinfo{person}{Torsten Hoefler}, \bibinfo{person}{Greg
  Bronevetsky}, \bibinfo{person}{Brian Barrett}, \bibinfo{person}{Bronis~R. de
  Supinski}, {and} \bibinfo{person}{Andrew Lumsdaine}.}
  \bibinfo{year}{2010}\natexlab{}.
\newblock \showarticletitle{Efficient MPI Support for Advanced Hybrid
  Programming Models}. In \bibinfo{booktitle}{\emph{Recent Advances in the
  Message Passing Interface}}, \bibfield{editor}{\bibinfo{person}{Rainer
  Keller}, \bibinfo{person}{Edgar Gabriel}, \bibinfo{person}{Michael Resch},
  {and} \bibinfo{person}{Jack Dongarra}} (Eds.). \bibinfo{publisher}{Springer
  Berlin Heidelberg}.
\newblock
\showISBNx{978-3-642-15646-5}


\bibitem[\protect\citeauthoryear{Hoefler, Di~Girolamo, Taranov, Grant, and
  Brightwell}{Hoefler et~al\mbox{.}}{2017}]%
        {Hoefler:2017:SHS}
\bibfield{author}{\bibinfo{person}{Torsten Hoefler}, \bibinfo{person}{Salvatore
  Di~Girolamo}, \bibinfo{person}{Konstantin Taranov}, \bibinfo{person}{Ryan~E.
  Grant}, {and} \bibinfo{person}{Ron Brightwell}.}
  \bibinfo{year}{2017}\natexlab{}.
\newblock \showarticletitle{sPIN: High-performance Streaming Processing In the
  Network}. In \bibinfo{booktitle}{\emph{Proceedings of the International
  Conference for High Performance Computing, Networking, Storage and Analysis}}
  \emph{(\bibinfo{series}{SC '17})}. \bibinfo{publisher}{ACM},
  \bibinfo{pages}{59:1--59:16}.
\newblock
\urldef\tempurl%
\url{https://doi.org/10.1145/3126908.3126970}
\showDOI{\tempurl}


\bibitem[\protect\citeauthoryear{IEEE and Group}{IEEE and Group}{2018}]%
        {POSIX2018}
\bibfield{author}{\bibinfo{person}{IEEE} {and} \bibinfo{person}{The~Open
  Group}.} \bibinfo{year}{2018}\natexlab{}.
\newblock \bibinfo{booktitle}{\emph{{The Open Group Base Specifications Issue
  7}}}.
\newblock \bibinfo{type}{IEEE Std} {1003.1-2017}. \bibinfo{institution}{IEEE}.
\newblock
\urldef\tempurl%
\url{https://pubs.opengroup.org/onlinepubs/9699919799/}
\showURL{%
\tempurl}
\newblock
\shownote{Last accessed April 24, 2020.}


\bibitem[\protect\citeauthoryear{Islam, Zheng, Sur, Langer, and Garzaran}{Islam
  et~al\mbox{.}}{2019}]%
        {Islam:2019:MUH}
\bibfield{author}{\bibinfo{person}{Nusrat~Sharmin Islam},
  \bibinfo{person}{Gengbin Zheng}, \bibinfo{person}{Sayantan Sur},
  \bibinfo{person}{Akhil Langer}, {and} \bibinfo{person}{Maria Garzaran}.}
  \bibinfo{year}{2019}\natexlab{}.
\newblock \showarticletitle{Minimizing the Usage of Hardware Counters for
  Collective Communication Using Triggered Operations}. In
  \bibinfo{booktitle}{\emph{Proceedings of the 26th European MPI Users’ Group
  Meeting}} (Z\"{u}rich, Switzerland) \emph{(\bibinfo{series}{EuroMPI ’19})}.
  \bibinfo{publisher}{Association for Computing Machinery}.
\newblock
\urldef\tempurl%
\url{https://doi.org/10.1145/3343211.3343222}
\showDOI{\tempurl}


\bibitem[\protect\citeauthoryear{{Iwasaki}, {Amer}, {Taura}, {Seo}, and
  {Balaji}}{{Iwasaki} et~al\mbox{.}}{2019}]%
        {Iwasaki:2019:BOLT}
\bibfield{author}{\bibinfo{person}{S. {Iwasaki}}, \bibinfo{person}{A. {Amer}},
  \bibinfo{person}{K. {Taura}}, \bibinfo{person}{S. {Seo}}, {and}
  \bibinfo{person}{P. {Balaji}}.} \bibinfo{year}{2019}\natexlab{}.
\newblock \showarticletitle{BOLT: Optimizing OpenMP Parallel Regions with
  User-Level Threads}. In \bibinfo{booktitle}{\emph{2019 28th International
  Conference on Parallel Architectures and Compilation Techniques (PACT)}}.
  \bibinfo{pages}{29--42}.
\newblock


\bibitem[\protect\citeauthoryear{Kaiser, Heller, Adelstein-Lelbach, Serio, and
  Fey}{Kaiser et~al\mbox{.}}{2014}]%
        {hpx:2014}
\bibfield{author}{\bibinfo{person}{Hartmut Kaiser}, \bibinfo{person}{Thomas
  Heller}, \bibinfo{person}{Bryce Adelstein-Lelbach}, \bibinfo{person}{Adrian
  Serio}, {and} \bibinfo{person}{Dietmar Fey}.}
  \bibinfo{year}{2014}\natexlab{}.
\newblock \showarticletitle{{HPX: A Task Based Programming Model in a Global
  Address Space}}. In \bibinfo{booktitle}{\emph{Proceedings of the 8th
  International Conference on Partitioned Global Address Space Programming
  Models}} \emph{(\bibinfo{series}{PGAS '14})}. \bibinfo{publisher}{ACM},
  \bibinfo{pages}{6:1--6:11}.
\newblock
\urldef\tempurl%
\url{https://doi.org/10.1145/2676870.2676883}
\showDOI{\tempurl}


\bibitem[\protect\citeauthoryear{Kowalke}{Kowalke}{2013}]%
        {BoostFiber}
\bibfield{author}{\bibinfo{person}{Oliver Kowalke}.}
  \bibinfo{year}{2013}\natexlab{}.
\newblock \bibinfo{booktitle}{\emph{Boost.Fiber -- Overview}}.
\newblock
\urldef\tempurl%
\url{https://www.boost.org/doc/libs/1_72_0/libs/fiber/doc/html/fiber/overview.html}
\showURL{%
\tempurl}
\newblock
\shownote{Last accessed April 24, 2020.}


\bibitem[\protect\citeauthoryear{Latham, Gropp, Ross, and Thakur}{Latham
  et~al\mbox{.}}{2007}]%
        {Latham:2007:EMG}
\bibfield{author}{\bibinfo{person}{Robert Latham}, \bibinfo{person}{William
  Gropp}, \bibinfo{person}{Robert Ross}, {and} \bibinfo{person}{Rajeev
  Thakur}.} \bibinfo{year}{2007}\natexlab{}.
\newblock \showarticletitle{Extending the MPI-2 Generalized Request Interface}.
  In \bibinfo{booktitle}{\emph{Proceedings of the 14th European Conference on
  Recent Advances in Parallel Virtual Machine and Message Passing Interface}}
  \emph{(\bibinfo{series}{PVM/MPI'07})}. \bibinfo{publisher}{Springer-Verlag}.
\newblock


\bibitem[\protect\citeauthoryear{Linux man-pages project}{Linux man-pages
  project}{2019a}]%
        {Makecontext:Linux}
Linux man-pages project \bibinfo{year}{2019}\natexlab{a}.
\newblock \bibinfo{booktitle}{\emph{makecontext, swapcontext -- manipulate user
  context}}.
\newblock Linux man-pages project.
\newblock
\urldef\tempurl%
\url{http://man7.org/linux/man-pages/man3/makecontext.3.html}
\showURL{%
\tempurl}
\newblock
\shownote{Last accessed April 24, 2020.}


\bibitem[\protect\citeauthoryear{Linux man-pages project}{Linux man-pages
  project}{2019b}]%
        {Pthread:Linux}
Linux man-pages project \bibinfo{year}{2019}\natexlab{b}.
\newblock \bibinfo{booktitle}{\emph{pthreads -- POSIX threads}}.
\newblock Linux man-pages project.
\newblock
\urldef\tempurl%
\url{http://man7.org/linux/man-pages/man7/pthreads.7.html}
\showURL{%
\tempurl}
\newblock
\shownote{Last accessed April 24, 2020.}


\bibitem[\protect\citeauthoryear{{Lu}, {Seo}, and {Balaji}}{{Lu}
  et~al\mbox{.}}{2015}]%
        {Lu:2015:MUO}
\bibfield{author}{\bibinfo{person}{H. {Lu}}, \bibinfo{person}{S. {Seo}}, {and}
  \bibinfo{person}{P. {Balaji}}.} \bibinfo{year}{2015}\natexlab{}.
\newblock \showarticletitle{MPI+ULT: Overlapping Communication and Computation
  with User-Level Threads}. In \bibinfo{booktitle}{\emph{2015 IEEE 17th
  International Conference on High Performance Computing and Communications,
  2015 IEEE 7th International Symposium on Cyberspace Safety and Security, and
  2015 IEEE 12th International Conference on Embedded Software and Systems}}.
  \bibinfo{pages}{444--454}.
\newblock
\showISSN{null}
\urldef\tempurl%
\url{https://doi.org/10.1109/HPCC-CSS-ICESS.2015.82}
\showDOI{\tempurl}


\bibitem[\protect\citeauthoryear{Mattsson}{Mattsson}{1996}]%
        {Mattsson:96:OOF}
\bibfield{author}{\bibinfo{person}{Michael Mattsson}.}
  \bibinfo{year}{1996}\natexlab{}.
\newblock \bibinfo{title}{Object-Oriented Frameworks - A survey of
  methodological issues}.
\newblock \bibinfo{howpublished}{Licentiate thesis}.
\newblock
\urldef\tempurl%
\url{http://citeseerx.ist.psu.edu/viewdoc/download?doi=10.1.1.36.1424&rep=rep1&type=pdf}
\showURL{%
\tempurl}


\bibitem[\protect\citeauthoryear{Mercier, Trahay, Buntinas, and Brunet}{Mercier
  et~al\mbox{.}}{2009}]%
        {Mercier:2009:NMA}
\bibfield{author}{\bibinfo{person}{Guillaume Mercier},
  \bibinfo{person}{François Trahay}, \bibinfo{person}{Darius Buntinas}, {and}
  \bibinfo{person}{Elisabeth Brunet}.} \bibinfo{year}{2009}\natexlab{}.
\newblock \showarticletitle{{NewMadeleine: An Efficient Support for
  High-Performance Networks in MPICH2}}. In
  \bibinfo{booktitle}{\emph{{Proceedings of the International Parallel and
  Distributed Processing Symposium (IPDPS)}}}.
\newblock


\bibitem[\protect\citeauthoryear{Microsoft}{Microsoft}{2018a}]%
        {MSThreads}
Microsoft \bibinfo{year}{2018}\natexlab{a}.
\newblock \bibinfo{booktitle}{\emph{About Processes and Threads}}.
\newblock Microsoft.
\newblock
\urldef\tempurl%
\url{https://docs.microsoft.com/en-us/windows/win32/procthread/about-processes-and-threads}
\showURL{%
\tempurl}
\newblock
\shownote{Last accessed April 24, 2020.}


\bibitem[\protect\citeauthoryear{Microsoft}{Microsoft}{2018b}]%
        {MSFibers}
Microsoft \bibinfo{year}{2018}\natexlab{b}.
\newblock \bibinfo{booktitle}{\emph{Fibers}}.
\newblock Microsoft.
\newblock
\urldef\tempurl%
\url{https://docs.microsoft.com/en-us/windows/win32/procthread/fibers}
\showURL{%
\tempurl}
\newblock
\shownote{Last accessed April 24, 2020.}


\bibitem[\protect\citeauthoryear{MPI Forum}{MPI v3.1}{2015}]%
        {mpi3.1}
MPI v3.1 \bibinfo{year}{2015}\natexlab{}.
\newblock \bibinfo{booktitle}{\emph{{MPI: A Message-Passing Interface
  Standard}}}.
\newblock \bibinfo{type}{{T}echnical {R}eport}.
\newblock
\urldef\tempurl%
\url{http://mpi-forum.org/docs/mpi-3.1/mpi31-report.pdf}
\showURL{%
\tempurl}
\newblock
\shownote{Last accessed April 24, 2020.}


\bibitem[\protect\citeauthoryear{{NASA Advanced Supercomputing Division}}{{NASA
  Advanced Supercomputing Division}}{[n.d.]}]%
        {NPB:2019:ProbSizes}
\bibfield{author}{\bibinfo{person}{{NASA Advanced Supercomputing Division}}.}
  \bibinfo{year}{[n.d.]}\natexlab{}.
\newblock \bibinfo{booktitle}{\emph{{Problem Sizes and Parameters in NAS
  Parallel Benchmarks}}}.
\newblock
\urldef\tempurl%
\url{https://www.nas.nasa.gov/publications/npb_problem_sizes.html}
\showURL{%
\tempurl}


\bibitem[\protect\citeauthoryear{{OpenFabrics Interfaces Working
  Group}}{{OpenFabrics Interfaces Working Group}}{2017}]%
        {OFI:2017}
\bibfield{author}{\bibinfo{person}{{OpenFabrics Interfaces Working Group}}.}
  \bibinfo{year}{2017}\natexlab{}.
\newblock \bibinfo{booktitle}{\emph{High Performance Network Programming with
  OFI}}.
\newblock
\urldef\tempurl%
\url{https://github.com/ofiwg/ofi-guide/blob/master/OFIGuide.md}
\showURL{%
\tempurl}
\newblock
\shownote{Last accessed May 14, 2020.}


\bibitem[\protect\citeauthoryear{OpenMP Architecture Review Board}{OpenMP
  Architecture Review Board}{2018}]%
        {openmp5.0}
OpenMP Architecture Review Board \bibinfo{year}{2018}\natexlab{}.
\newblock \bibinfo{booktitle}{\emph{{OpenMP Application Programming Interface,
  Version 5.0}}}.
\newblock OpenMP Architecture Review Board.
\newblock
\urldef\tempurl%
\url{https://www.openmp.org/wp-content/uploads/OpenMP-API-Specification-5.0.pdf}
\showURL{%
\tempurl}
\newblock
\shownote{Last accessed April 24, 2020.}


\bibitem[\protect\citeauthoryear{{Patinyasakdikul}, {Eberius}, {Bosilca}, and
  {Hjelm}}{{Patinyasakdikul} et~al\mbox{.}}{2019}]%
        {Patinyasakdikul:2019:GMT}
\bibfield{author}{\bibinfo{person}{T. {Patinyasakdikul}}, \bibinfo{person}{D.
  {Eberius}}, \bibinfo{person}{G. {Bosilca}}, {and} \bibinfo{person}{N.
  {Hjelm}}.} \bibinfo{year}{2019}\natexlab{}.
\newblock \showarticletitle{Give MPI Threading a Fair Chance: A Study of
  Multithreaded MPI Designs}. In \bibinfo{booktitle}{\emph{2019 IEEE
  International Conference on Cluster Computing (CLUSTER)}}.
  \bibinfo{pages}{1--11}.
\newblock
\urldef\tempurl%
\url{https://doi.org/10.1109/CLUSTER.2019.8891015}
\showDOI{\tempurl}


\bibitem[\protect\citeauthoryear{Pritchard, Gorodetsky, and Buntinas}{Pritchard
  et~al\mbox{.}}{2011}]%
        {Pritchard:2011:uGNI}
\bibfield{author}{\bibinfo{person}{Howard Pritchard}, \bibinfo{person}{Igor
  Gorodetsky}, {and} \bibinfo{person}{Darius Buntinas}.}
  \bibinfo{year}{2011}\natexlab{}.
\newblock \showarticletitle{A uGNI-Based MPICH2 Nemesis Network Module for the
  Cray XE}. In \bibinfo{booktitle}{\emph{Recent Advances in the Message Passing
  Interface}}, \bibfield{editor}{\bibinfo{person}{Yiannis Cotronis},
  \bibinfo{person}{Anthony Danalis}, \bibinfo{person}{Dimitrios~S.
  Nikolopoulos}, {and} \bibinfo{person}{Jack Dongarra}} (Eds.).
  \bibinfo{publisher}{Springer Berlin Heidelberg}.
\newblock


\bibitem[\protect\citeauthoryear{Reinders}{Reinders}{2007}]%
        {Reinders:2007:TBB}
\bibfield{author}{\bibinfo{person}{James Reinders}.}
  \bibinfo{year}{2007}\natexlab{}.
\newblock \bibinfo{booktitle}{\emph{{Intel threading building blocks:
  outfitting C++ for multi-core processor parallelism}}}.
\newblock \bibinfo{publisher}{O'Reilly \& Associates}.
\newblock


\bibitem[\protect\citeauthoryear{Reynolds}{Reynolds}{1993}]%
        {Reynolds;1993:TDC}
\bibfield{author}{\bibinfo{person}{John~C. Reynolds}.}
  \bibinfo{year}{1993}\natexlab{}.
\newblock \showarticletitle{The discoveries of continuations}.
\newblock \bibinfo{journal}{\emph{LISP and Symbolic Computation}}
  (\bibinfo{date}{Nov.} \bibinfo{year}{1993}).
\newblock
\urldef\tempurl%
\url{https://doi.org/10.1007/BF01019459}
\showDOI{\tempurl}


\bibitem[\protect\citeauthoryear{Sala, Bell\'{o}n, Farr\'{e}, Teruel, Perez,
  Pe\~{n}a, Holmes, Beltran, and Labarta}{Sala et~al\mbox{.}}{2018}]%
        {Sala:2018:IIM}
\bibfield{author}{\bibinfo{person}{Kevin Sala}, \bibinfo{person}{Jorge
  Bell\'{o}n}, \bibinfo{person}{Pau Farr\'{e}}, \bibinfo{person}{Xavier
  Teruel}, \bibinfo{person}{Josep~M. Perez}, \bibinfo{person}{Antonio~J.
  Pe\~{n}a}, \bibinfo{person}{Daniel Holmes}, \bibinfo{person}{Vicen\c{c}
  Beltran}, {and} \bibinfo{person}{Jesus Labarta}.}
  \bibinfo{year}{2018}\natexlab{}.
\newblock \showarticletitle{Improving the Interoperability between MPI and
  Task-Based Programming Models}. In \bibinfo{booktitle}{\emph{Proceedings of
  the 25th European MPI Users’ Group Meeting}}
  \emph{(\bibinfo{series}{EuroMPI’18})}. \bibinfo{publisher}{Association for
  Computing Machinery}, \bibinfo{address}{New York, NY, USA}.
\newblock
\urldef\tempurl%
\url{https://doi.org/10.1145/3236367.3236382}
\showDOI{\tempurl}


\bibitem[\protect\citeauthoryear{Sala, Teruel, Perez, Peña, Beltran, and
  Labarta}{Sala et~al\mbox{.}}{2019}]%
        {Sala:2019:TAMPI}
\bibfield{author}{\bibinfo{person}{Kevin Sala}, \bibinfo{person}{Xavier
  Teruel}, \bibinfo{person}{Josep~M. Perez}, \bibinfo{person}{Antonio~J.
  Peña}, \bibinfo{person}{Vicenç Beltran}, {and} \bibinfo{person}{Jesus
  Labarta}.} \bibinfo{year}{2019}\natexlab{}.
\newblock \showarticletitle{Integrating blocking and non-blocking MPI
  primitives with task-based programming models}.
\newblock \bibinfo{journal}{\emph{Parallel Comput.}}  \bibinfo{volume}{85}
  (\bibinfo{date}{Jul} \bibinfo{year}{2019}), \bibinfo{pages}{153–166}.
\newblock
\showISSN{0167-8191}
\urldef\tempurl%
\url{https://doi.org/10.1016/j.parco.2018.12.008}
\showDOI{\tempurl}


\bibitem[\protect\citeauthoryear{Schonbein, Grant, Dosanjh, and
  Arnold}{Schonbein et~al\mbox{.}}{2019}]%
        {Schonbein:2019:INCA}
\bibfield{author}{\bibinfo{person}{Whit Schonbein}, \bibinfo{person}{Ryan~E.
  Grant}, \bibinfo{person}{Matthew G.~F. Dosanjh}, {and}
  \bibinfo{person}{Dorian Arnold}.} \bibinfo{year}{2019}\natexlab{}.
\newblock \showarticletitle{INCA: In-Network Compute Assistance}. In
  \bibinfo{booktitle}{\emph{Proceedings of the International Conference for
  High Performance Computing, Networking, Storage and Analysis}}
  \emph{(\bibinfo{series}{SC ’19})}. \bibinfo{publisher}{Association for
  Computing Machinery}.
\newblock
\urldef\tempurl%
\url{https://doi.org/10.1145/3295500.3356153}
\showDOI{\tempurl}


\bibitem[\protect\citeauthoryear{{Schuchart}, {Bouteiller}, and
  {Bosilca}}{{Schuchart} et~al\mbox{.}}{2019}]%
        {Schuchart:2019:UMR}
\bibfield{author}{\bibinfo{person}{J. {Schuchart}}, \bibinfo{person}{A.
  {Bouteiller}}, {and} \bibinfo{person}{G. {Bosilca}}.}
  \bibinfo{year}{2019}\natexlab{}.
\newblock \showarticletitle{{Using MPI-3 RMA for Active Messages}}. In
  \bibinfo{booktitle}{\emph{2019 IEEE/ACM Workshop on Exascale MPI (ExaMPI)}}.
  \bibinfo{pages}{47--56}.
\newblock


\bibitem[\protect\citeauthoryear{Schuchart and Gracia}{Schuchart and
  Gracia}{2019}]%
        {Schuchart:2019:GTD}
\bibfield{author}{\bibinfo{person}{Joseph Schuchart} {and}
  \bibinfo{person}{Jos{\'e} Gracia}.} \bibinfo{year}{2019}\natexlab{}.
\newblock \showarticletitle{{Global Task Data-Dependencies in PGAS
  Applications}}. In \bibinfo{booktitle}{\emph{High Performance Computing}},
  \bibfield{editor}{\bibinfo{person}{Mich{\`e}le Weiland},
  \bibinfo{person}{Guido Juckeland}, \bibinfo{person}{Carsten Trinitis}, {and}
  \bibinfo{person}{Ponnuswamy Sadayappan}} (Eds.). \bibinfo{publisher}{Springer
  International Publishing}.
\newblock


\bibitem[\protect\citeauthoryear{Schuchart, Tsugane, Gracia, and
  Sato}{Schuchart et~al\mbox{.}}{2018}]%
        {Schuchart:2018:TIT}
\bibfield{author}{\bibinfo{person}{Joseph Schuchart}, \bibinfo{person}{Keisuke
  Tsugane}, \bibinfo{person}{Jos{\'e} Gracia}, {and} \bibinfo{person}{Mitsuhisa
  Sato}.} \bibinfo{year}{2018}\natexlab{}.
\newblock \showarticletitle{The Impact of Taskyield on the Design of Tasks
  Communicating Through MPI}. In \bibinfo{booktitle}{\emph{Evolving OpenMP for
  Evolving Architectures}}, \bibfield{editor}{\bibinfo{person}{Bronis~R.
  de~Supinski}, \bibinfo{person}{Pedro Valero-Lara}, \bibinfo{person}{Xavier
  Martorell}, \bibinfo{person}{Sergi Mateo~Bellido}, {and}
  \bibinfo{person}{Jesus Labarta}} (Eds.). \bibinfo{publisher}{Springer
  International Publishing}, \bibinfo{pages}{3--17}.
\newblock
\urldef\tempurl%
\url{https://doi.org/10.1007/978-3-319-98521-3_1}
\showDOI{\tempurl}
\newblock
\shownote{Awarded Best Paper.}


\bibitem[\protect\citeauthoryear{{Seo}, {Amer}, {Balaji}, {Bordage}, {Bosilca},
  {Brooks}, {Carns}, {Castelló}, {Genet}, {Herault}, {Iwasaki}, {Jindal},
  {Kalé}, {Krishnamoorthy}, {Lifflander}, {Lu}, {Meneses}, {Snir}, {Sun},
  {Taura}, and {Beckman}}{{Seo} et~al\mbox{.}}{2018}]%
        {Seo:2018:AAL}
\bibfield{author}{\bibinfo{person}{S. {Seo}}, \bibinfo{person}{A. {Amer}},
  \bibinfo{person}{P. {Balaji}}, \bibinfo{person}{C. {Bordage}},
  \bibinfo{person}{G. {Bosilca}}, \bibinfo{person}{A. {Brooks}},
  \bibinfo{person}{P. {Carns}}, \bibinfo{person}{A. {Castelló}},
  \bibinfo{person}{D. {Genet}}, \bibinfo{person}{T. {Herault}},
  \bibinfo{person}{S. {Iwasaki}}, \bibinfo{person}{P. {Jindal}},
  \bibinfo{person}{L.~V. {Kalé}}, \bibinfo{person}{S. {Krishnamoorthy}},
  \bibinfo{person}{J. {Lifflander}}, \bibinfo{person}{H. {Lu}},
  \bibinfo{person}{E. {Meneses}}, \bibinfo{person}{M. {Snir}},
  \bibinfo{person}{Y. {Sun}}, \bibinfo{person}{K. {Taura}}, {and}
  \bibinfo{person}{P. {Beckman}}.} \bibinfo{year}{2018}\natexlab{}.
\newblock \showarticletitle{{Argobots: A Lightweight Low-Level Threading and
  Tasking Framework}}.
\newblock \bibinfo{journal}{\emph{IEEE Transactions on Parallel and Distributed
  Systems}} \bibinfo{volume}{29}, \bibinfo{number}{3} (\bibinfo{date}{March}
  \bibinfo{year}{2018}), \bibinfo{pages}{512--526}.
\newblock
\urldef\tempurl%
\url{https://doi.org/10.1109/TPDS.2017.2766062}
\showDOI{\tempurl}


\bibitem[\protect\citeauthoryear{Silberschatz, Galvin, and Gagne}{Silberschatz
  et~al\mbox{.}}{2014}]%
        {Silberschatz:2014:OSC}
\bibfield{author}{\bibinfo{person}{Abraham Silberschatz},
  \bibinfo{person}{Peter~B. Galvin}, {and} \bibinfo{person}{Greg Gagne}.}
  \bibinfo{year}{2014}\natexlab{}.
\newblock \bibinfo{booktitle}{\emph{Operating System Concepts}
  (\bibinfo{edition}{9} ed.)}.
\newblock \bibinfo{publisher}{John Wiley and Sons}.
\newblock
\showISBNx{978-11180-9375-7}


\bibitem[\protect\citeauthoryear{Stark, Barrett, Grant, Olivier, Pedretti, and
  Vaughan}{Stark et~al\mbox{.}}{2014}]%
        {Stark:2014:EECS}
\bibfield{author}{\bibinfo{person}{Dylan~T. Stark}, \bibinfo{person}{Richard~F.
  Barrett}, \bibinfo{person}{Ryan~E. Grant}, \bibinfo{person}{Stephen~L.
  Olivier}, \bibinfo{person}{Kevin~T. Pedretti}, {and}
  \bibinfo{person}{Courtenay~T. Vaughan}.} \bibinfo{year}{2014}\natexlab{}.
\newblock \showarticletitle{Early Experiences Co-Scheduling Work and
  Communication Tasks for Hybrid MPI+X Applications}. In
  \bibinfo{booktitle}{\emph{Proceedings of the 2014 Workshop on Exascale MPI}}
  \emph{(\bibinfo{series}{ExaMPI ’14})}. \bibinfo{publisher}{IEEE Press}.
\newblock
\urldef\tempurl%
\url{https://doi.org/10.1109/ExaMPI.2014.6}
\showDOI{\tempurl}


\bibitem[\protect\citeauthoryear{Sun Microsystems, Inc}{Sun Microsystems,
  Inc}{2002}]%
        {SolarisMT}
Sun Microsystems, Inc \bibinfo{year}{2002}\natexlab{}.
\newblock \bibinfo{booktitle}{\emph{Multithreading in the Solaris Operating
  Environment}}.
\newblock Sun Microsystems, Inc.
\newblock
\urldef\tempurl%
\url{https://web.archive.org/web/20090226174929/http://www.sun.com/software/whitepapers/solaris9/multithread.pdf}
\showURL{%
\tempurl}
\newblock
\shownote{Last accessed April 24, 2020.}


\bibitem[\protect\citeauthoryear{The FreeBSD Project}{The FreeBSD
  Project}{2018}]%
        {Pthread:FreeBDS}
The FreeBSD Project \bibinfo{year}{2018}\natexlab{}.
\newblock \bibinfo{booktitle}{\emph{pthread -- POSIX thread functions}}.
\newblock The FreeBSD Project.
\newblock
\urldef\tempurl%
\url{https://www.freebsd.org/cgi/man.cgi?query=pthread}
\showURL{%
\tempurl}
\newblock
\shownote{Last accessed April 24, 2020.}


\bibitem[\protect\citeauthoryear{The NetBSD project}{The NetBSD
  project}{2016}]%
        {Pthread:NetBSD}
The NetBSD project \bibinfo{year}{2016}\natexlab{}.
\newblock \bibinfo{booktitle}{\emph{pthread -- POSIX Threads Library}}.
\newblock The NetBSD project.
\newblock
\urldef\tempurl%
\url{https://netbsd.gw.com/cgi-bin/man-cgi?pthread+3.i386+NetBSD-8.0}
\showURL{%
\tempurl}
\newblock
\shownote{Last accessed April 24, 2020.}


\bibitem[\protect\citeauthoryear{Unangst, Miller, Hyman, Moerbeek, and
  Guenther}{Unangst et~al\mbox{.}}{2019}]%
        {Pthread:OpenBSD}
\bibfield{author}{\bibinfo{person}{Ted Unangst}, \bibinfo{person}{Kurt Miller},
  \bibinfo{person}{Marco~S Hyman}, \bibinfo{person}{Otto Moerbeek}, {and}
  \bibinfo{person}{Philip Guenther}.} \bibinfo{year}{2019}\natexlab{}.
\newblock \bibinfo{booktitle}{\emph{pthreads -- POSIX 1003.1c thread
  interface}}.
\newblock The OpenBSD project.
\newblock
\urldef\tempurl%
\url{https://man.openbsd.org/pthreads.3}
\showURL{%
\tempurl}
\newblock
\shownote{Last accessed April 24, 2020.}


\bibitem[\protect\citeauthoryear{{Unified Communication Framework
  Consortium}}{{Unified Communication Framework Consortium}}{2019}]%
        {UCX:1.6}
\bibfield{author}{\bibinfo{person}{{Unified Communication Framework
  Consortium}}.} \bibinfo{year}{2019}\natexlab{}.
\newblock \bibinfo{booktitle}{\emph{{UCX: Unified Communication X API Standard
  v1.6}}}.
\newblock Unified Communication Framework Consortium.
\newblock
\urldef\tempurl%
\url{https://github.com/openucx/ucx/wiki/api-doc/v1.6/ucx-v1.6.pdf}
\showURL{%
\tempurl}


\bibitem[\protect\citeauthoryear{Vahalia}{Vahalia}{1998}]%
        {UNIXInternals}
\bibfield{author}{\bibinfo{person}{Uresh Vahalia}.}
  \bibinfo{year}{1998}\natexlab{}.
\newblock \bibinfo{booktitle}{\emph{UNIX Internals: The New Frontiers}}.
\newblock \bibinfo{publisher}{Pearson}.
\newblock
\showISBNx{978-0131019089}


\bibitem[\protect\citeauthoryear{Wasi-ur Rahman, Ozog, and Dinan}{Wasi-ur
  Rahman et~al\mbox{.}}{2020}]%
        {Rahman:SCO:2020}
\bibfield{author}{\bibinfo{person}{Md. Wasi-ur Rahman}, \bibinfo{person}{David
  Ozog}, {and} \bibinfo{person}{James Dinan}.} \bibinfo{year}{2020}\natexlab{}.
\newblock \showarticletitle{Simplifying Communication Overlap in OpenSHMEM
  Through Integrated User-Level Thread Scheduling}. In
  \bibinfo{booktitle}{\emph{High Performance Computing}},
  \bibfield{editor}{\bibinfo{person}{Ponnuswamy Sadayappan},
  \bibinfo{person}{Bradford~L. Chamberlain}, \bibinfo{person}{Guido Juckeland},
  {and} \bibinfo{person}{Hatem Ltaief}} (Eds.). \bibinfo{publisher}{Springer
  International Publishing}.
\newblock


\bibitem[\protect\citeauthoryear{{Wheeler}, {Murphy}, and {Thain}}{{Wheeler}
  et~al\mbox{.}}{2008}]%
        {Wheeler:2008:QAA}
\bibfield{author}{\bibinfo{person}{K.~B. {Wheeler}}, \bibinfo{person}{R.~C.
  {Murphy}}, {and} \bibinfo{person}{D. {Thain}}.}
  \bibinfo{year}{2008}\natexlab{}.
\newblock \showarticletitle{Qthreads: An API for programming with millions of
  lightweight threads}. In \bibinfo{booktitle}{\emph{2008 IEEE International
  Symposium on Parallel and Distributed Processing}}.
\newblock
\showISSN{1530-2075}
\urldef\tempurl%
\url{https://doi.org/10.1109/IPDPS.2008.4536359}
\showDOI{\tempurl}


\end{thebibliography}

\end{document}